\documentclass[12pt]{article}
\usepackage[utf8]{inputenc}
\usepackage{geometry}
\usepackage{graphicx}
\usepackage{array}
\usepackage{subfig}
\usepackage{slashed}
\usepackage{amsmath}
\usepackage{amsfonts}
\usepackage{amssymb}
\usepackage[cm]{fullpage}
\usepackage{cite}
\usepackage{epstopdf}
\usepackage{comment}
\usepackage{hyperref}
\usepackage{cancel}
\usepackage{mathrsfs}

\usepackage{cleveref}
\crefname{section}{§}{§§}
\Crefname{section}{§}{§§}

\usepackage{booktabs} \newcommand{\ra}[1]{\renewcommand{\arraystretch}{#1}}

\numberwithin{equation}{section}

\def\p{\partial}

\def\0{{(0)}}
\def\1{{(1)}}
\def\2{{(2)}}

\def\<{\langle }
\def\>{\rangle }

\newcommand{\bea}{\begin{eqnarray}}

\newcommand{\eea}{\end{eqnarray}}

\newcommand{\be}{\begin{equation}}
\newcommand{\ee}{\end{equation}}
\newcommand{\ba}{\begin{align}}
\newcommand{\ea}{\end{align}}

\newcommand{\cc}[2]{c\genfrac{[}{]}{0pt}{}{#1}{#2}}

   \makeatletter
  \let\over=\@@over \let\overwithdelims=\@@overwithdelims
  \let\atop=\@@atop \let\atopwithdelims=\@@atopwithdelims
  \let\above=\@@above \let\abovewithdelims=\@@abovewithdelims
\renewcommand\section{\@startsection {section}{1}{\z@}%
                                   {-3.5ex \@plus -1ex \@minus -.2ex}
                                   {2.3ex \@plus.2ex}%
                                   {\normalfont\large\bfseries}}

\renewcommand\subsection{\@startsection{subsection}{2}{\z@}%
                                     {-3.25ex\@plus -1ex \@minus -.2ex}%
                                     {1.5ex \@plus .2ex}%
                                     {\normalfont\bfseries}}

\newcommand{\beq}{\begin{equation}}
\newcommand{\eeq}{\end{equation}}
\newcommand{\beqa}{\begin{eqnarray}}
\newcommand{\eeqa}{\end{eqnarray}}
\newcommand{\beqar}{\begin{eqnarray*}}

\def\[{\[}
\def\]{\]}

\newcommand{\bd}[1]{\begin{fmffile}{#1}\begin{fmfgraph*}}
\newcommand{\ed}{\end{fmfgraph*}\end{fmffile}}

\begin{document}

\begin{titlepage}

\begin{flushright}
	CERN-PH-TH/2016-179
\end{flushright}

\unitlength = 1mm~\\
\vskip 1cm
\begin{center}

{\LARGE{\textsc{Chiral Heterotic Strings  \\[0.3cm] with Positive Cosmological Constant}}}

\vspace{0.8cm}
Ioannis Florakis\,{}\footnote{{\tt florakis@cern.ch}, After $1^{\rm rst}$ September 2016 at LPTHE, 4 Place Jussieu, Paris, France} and John Rizos\,{}\footnote{\tt irizos@uoi.gr}

\vspace{1cm}

{\it  ${}^1$ Theory Division - CERN, CH-1211 Geneva 23, Switzerland \\
	 ${}^2$ Department of Physics, University of Ioannina, GR45110 Ioannina, Greece 

}

\vspace{0.8cm}

\begin{abstract}
We present explicit examples of semi-realistic heterotic models with spontaneously broken supersymmetry, which dynamically lead to breaking scales much smaller than $M_{\rm Planck}$ and exponentially small positive values for the cosmological constant. Contrary to field theoretic intuition, we find that the global structure of the effective potential is significantly affected by contributions of massive and non-level matched string states and we investigate the conditions that dynamically ensure a number of desired properties.

\end{abstract}

\setcounter{footnote}{0}

\vspace{1.0cm}

\end{center}

\end{titlepage}

\pagestyle{empty}
\pagestyle{plain}

\def\vx{{\vec x}}
\def\p{\partial}
\def\po{$\cal P_O$}

\pagenumbering{arabic}

\tableofcontents
\bibliographystyle{utphys}

\section{Introduction}

Whenever supersymmetry is spontaneously broken in perturbative string theory, one is eventually called to face at least two fundamental issues. The first one relates to the possibility of encountering tachyonic modes in the physical string spectrum, signalling a tree level instability of the theory. These are level-matched modes with non-trivial winding which may become tachyonic in regions of moduli space sufficiently close to the string scale. This problem is essentially linked to the exponentially growing degeneracy of states of string theory and, for example, manifests itself as the Hagedorn instability in a thermal setup \cite{Atick:1988si}. The second one is related to the presence of a one-loop tadpole back-reacting on the classical vacuum \cite{Fischler:1986ci,Fischler:1986tb}.

The way to break supersymmetry spontaneously in closed string theory, that still admits a fully-fledged perturbative worldsheet description is the stringy version \cite{Rohm:1983aq,Kounnas:1988ye,Ferrara:1988jx,Kounnas:1989dk} of the Scherk-Schwarz mechanism \cite{Scherk:1978ta,Scherk:1979zr}. From a field theory perspective, it corresponds to a flat gauging of supergravity, generating a scalar potential at tree level with vanishing cosmological constant at its minimum. It is precisely at this point that  string theory may be exactly quantised and corresponds to the worldsheet CFT of a freely acting orbifold.

The Scherk-Schwarz mechanism is essentially a deformation of the theory by a symmetry operator $Q$, introducing a non-trivial monodromy to fields or vertex operators of the theory as one encircles a compact cycle of the internal manifold, $\Phi(x_\mu,y+2\pi R)=e^{iQ}\,\Phi(x_\mu,y)$. This induces a shift in the Kaluza-Klein spectrum of states charged under $Q$ and, in particular, gives rise to a mass gap inversely proportional to the radius of the compact cycle. In the framework of the supersymmetric heterotic string, the Scherk-Schwarz deformation upon identifying the generator $Q$ with the spacetime fermion number amounts to assigning different boundary conditions to bosonic and fermionic states within the same multiplet, and is responsible for the spontaneous breaking of supersymmetry with the breaking scale tied to the size of the compact dimension, $m_{3/2}\sim 1/R$.

In the formalism of gauged supergravity, Scherk-Schwarz corresponds to a special flat gauging inducing a non-trivial mass term for the gravitino as well as a tree-level scalar potential of the no-scale type \cite{Cremmer:1983bf}. Namely, as soon as one minimises the scalar potential with respect to the charged fields, the potential vanishes and the scalars neutral with respect to the gauging remain massless at tree level. These no-scale moduli precisely enter into the gravitino mass term, expanded around the minimum, and one obtains a family of vacua with the scale of supersymmetry breaking $m_{3/2}$ remaining undetermined at tree level. This no-scale structure is, of course, consistent with the fact that the moduli entering the one-loop partition function of the string correspond to consistent marginal deformations of the worldsheet CFT.

This situation changes drastically at the loop level, where the scalar potential receives radiative corrections, that may stabilise or even destabilise the no-scale moduli and opens the possibility for a dynamical determination of the supersymmetry breaking scale. In general, the effective potential at one loop as a function of the no-scale moduli $t_I$ is obtained by integrating the string partition function $Z(\tau_1,\tau_2;t_I)$ over the moduli space of the worldsheet torus $\Sigma_1$
\begin{align}
	V_{\rm one-loop}(t_I) = - \frac{1}{2(2\pi)^4} \int_{\mathcal F}\frac{d^2\tau}{\tau_2^3}\,Z(\tau,\bar\tau; t_I)\,,
	\label{potentialdef}
\end{align}
where $\tau=\tau_1+i\tau_2$ is the complex structure on $\Sigma_1$ and $\mathcal F = {\rm SL}(2;\mathbb Z)\backslash\mathbb{H}^+$ is a fundamental domain obtained as the quotient of the Teichm\"uller space by the mapping class group ${\rm SL}(2;\mathbb Z)$. We work in string units $\alpha'=1$.

Although non-supersymmetric closed string theories have been considered in many cases in the literature \cite{AlvarezGaume:1986jb,Dixon:1986iz,Ginsparg:1986wr,Nair:1986zn,Itoyama:1986ei,Taylor:1987uv,Toon:1990ij,Dienes:1995bx,Sasada:1995wq,Blum:1997cs,Blum:1997gw,Harvey:1998rc,Ghilencea:2001bv,Font:2002pq,Angelantonj:2006ut,Faraggi:2007tj,GatoRivera:2007yi,GatoRivera:2008zn,Florakis:2009sm,Faraggi:2009xy,Florakis:2010ty}, including also recent works in the context of string phenomenology \cite{Blaszczyk:2014qoa,Abel:2015oxa,Lukas:2015kca,Ashfaque:2015vta,Blaszczyk:2015zta,Nibbelink:2015vha}, the behaviour of string theories with broken supersymmetry still remains a largely unexplored terrain. Aside from controlling the spectrum of such theories, several open questions remain to be addressed. For instance, only very recently has a more systematic study of non-supersymmetric string gauge couplings been considered at the loop level \cite{Angelantonj:2014dia,Florakis:2015txa,Faraggi:2014eoa,Kounnas:2016gmz}. In most cases, attempts to make precise studies of string interactions in the absence of the benefits of supersymmetry, is quickly transformed into a cumbersome mathematical problem of taming quantum corrections to various couplings receiving contributions from the entire ensemble of Kaluza-Klein, winding, and string oscillators. Such couplings no longer display the special holomorphy properties enjoyed by their BPS counterparts in supersymmetric theories, and remain largely intractable. A notable exception is the remarkable universality discovered recently \cite{Angelantonj:2014dia,Florakis:2015txa} in the difference of gauge thresholds, which under certain specific conditions \cite{Angelantonj:2015nfa,Angelantonj:2016ibb}, guarantees the explicit solvability of the corresponding amplitudes.

Regardless of the inherent difficulties arising when dealing with non-supersymmetric string theory, considerable progress has been made recently in studying various one-loop amplitudes. One such example, is the construction of  \emph{super no-scale models} \cite{Harvey:1998rc,Angelantonj:1999gm,Shiu:1998he,Abel:2015oxa,Kounnas:2016gmz}. These are non-supersymmetric heterotic strings with Bose-Fermi degeneracy at the massless level, yielding at most exponentially small values for the vacuum amplitude, which appears to be a necessary ingredient in scenarios with a low scale for supersymmetry breaking. Aside from ensuring small values for the vacuum energy in the limit of large volume in the Scherk-Schwarz radius, such models have the additional virtue of softening the back-reaction problem.

We wish to stress, however, that, although necessary for suppressing the value of the cosmological constant at large volume of the internal space, the super no-scale requirement of a degeneracy  in the number of bosons and fermions $n_B=n_F$ in the massless sector is not sufficient to guarantee the positivity of the one-loop potential. In fact, as we discuss in this work, the global structure of the effective potential is crucially dependent on the behaviour of massive and even non level-matched string states around special self-dual points in moduli space.

The purpose of this paper is to investigate the rich structure of the one-loop effective potential as a function of the no-scale moduli, in the context of heterotic string theory with $\mathcal N=1\to 0$ spontaneously broken supersymmetry, while retaining some minimal phenomenological requirements, such as the presence of chiral matter. Obtaining an exact analytical grasp of the behaviour of the one loop effective potential as a function of the no-scale moduli throughout moduli space is a notoriously hard problem in string theory, mainly due to the absence of the simple holomorphy properties in the Teichm\"uller parameter $\tau$ enjoyed by BPS couplings, the complexity of the integrand as a function of several moduli $t_I$, as well as convergence issues arising from the exponential growth of the degeneracy of states of string theory. These impinge on the validity of standard unfolding techniques when applied to the evaluation of non-BPS protected amplitudes around self-dual points. Although powerful new techniques \cite{Angelantonj:2011br,Angelantonj:2012gw,Florakis:2013ura,Angelantonj:2013eja,Angelantonj:2015rxa,Florakis:2016boz} have been recently proposed for the evaluation of Schwinger-like integrals at one and higher genera, which are precisely tailored to capture the behaviour around self-dual points, their efficiency is mainly restricted to BPS protected quantities.

Nevertheless, it is possible to make some generic statements about the structure of the effective one-loop potential. On the one hand, self-dual points of the compactification lattice manifest themselves as extrema of the effective potential. On the other hand, the spontaneous nature of the supersymmetry breaking implies that supersymmetry be effectively recovered in the limit of infinite volume in the no-scale parameters entering the gravitino mass term, so that one expects the potential to vanish in these limits.

In the simplest heterotic models with Scherk-Schwarz breaking, \emph{e.g.} corresponding to the freely acting orbifold 
\begin{align}
	\mathbb Z_2 \ : \ g=(-1)^{F_{\rm s.t.}}\,\delta\,,
\end{align}
one mods out by the spacetime fermion number $F_{\rm s.t.}$, coupled to an order two shift $\delta$ along a circle $S^1$ of the internal manifold, $y\to y+\pi R$. In this case, the massless spectrum at generic radius is populated entirely by bosons, which in turn implies a negative value for the effective one-loop potential at large radii  $R\gg 1$, as computed in \cite{Antoniadis:1990ew}
\begin{align}
	V_{\rm one-loop}(t_I) \sim (n_F-n_B)/R^4\,.
\end{align}
Because of duality under $R\to 1/R$, one may naturally anticipate the shape of the potential to have the form of a ``puddle", stabilising the no-scale modulus $R$ and, hence, also the scale of supersymmetry breaking, to values of the order of the string scale. Of course, in this case, the value of the cosmological constant is both huge (in absolute value) and negative. Regardless of this issue, and even though this possibility may not be particularly attractive for phenomenology, e.g. for addressing the hierarchy problem, it does possess at least some positive features. For instance, one-loop corrections to the running of gauge couplings are finite and the decompactification problem \cite{Kiritsis:1996xd,Faraggi:2014eoa} does not arise in this case. 

On the other hand, in the absence of supersymmetry, the stabilisation of the radius around self-dual points could give rise to much more serious issues, related to the possibility of exciting tachyonic modes along some other directions of the classical moduli space, leading to the divergence of the one-loop potential. Indeed, the richness of the parameter space of no-scale moduli typically allows for tachyonic states to appear in the heterotic string spectrum as long as the supersymmetry breaking scale is of the order of the string scale. In that case, solving a strong backreaction problem associated with the condensation of the tachyonic modes to their true vacuum and re-quantizing the theory around it appears to be impossible to avoid \cite{Antoniadis:1991kh,Antoniadis:1999gz}. A full stability analysis \cite{Kutasov:1990sv,Dienes:1994np,Angelantonj:2006ut,Angelantonj:2008fz,Angelantonj:2010ic}, \cite{Florakis:2010ty} is therefore required, potentially necessitating the introduction of additional mechanisms, in order to secure the stability of the classical vacuum and the validity of perturbation theory. 

The situation might considerably improve if the dynamics of the problem were instead to attract the radius away from the string scale, \emph{i.e.} to regimes of large volume $R\gg 1$, in which no tachyonic modes can appear. This scenario opens the possibility of having supersymmetry broken at lower scales $m_{3/2}\ll 1$, while at the same time suppressing the value of the (now positive) cosmological constant. In this work we investigate this possibility in the context of heterotic theories with spontaneously broken $\mathcal N=1\to 0$ supersymmetry, admitting chiral matter and with an observable gauge group relevant for ${\rm SO}(10)$ GUT model building. We show that the space of models possessing these attractive features is not empty and we present explicit such examples which furnish a natural mechanism for dynamically protecting the theory against the development of tachyonic instabilities. 

A second related question that we address in this work, concerns the conditions for constructing heterotic theories with this desired behaviour in their one-loop potential. One might expect that a natural way to achieve this would be to require, in addition to the super no-scale requirement $n_B=n_F$ at the generic point in moduli space \cite{Abel:2015oxa,Kounnas:2016gmz}, that the massless spectrum be no longer dominated by bosons at the self-dual points, $n_B<n_F$. Perhaps not surprisingly, this simple requirement is not what determines the actual shape of the effective potential. As we shall see, the physics around points of enhanced symmetry is much more subtle, and the crucial role in determining the global morphology of the effective potential is actually played also by massive, as well as non-level matched states.

The paper is structured as follows. In section \ref{FFFscan} we outline our starting point for a computer-aided scan of models satisfying certain minimal phenomenological criteria and discuss the basic results. In section \ref{examplemodel} we set the scene for the analyses to follow by discussing in some detail a particular chiral, toy-model example and study  the form of its effective one-loop potential. In section \ref{counterexamplemodel} we move on to present a super no-scale counter-example model with exponentially suppressed value for the cosmological constant  which, however, does not lead to the desired form of the potential due to non-trivial effects arising around extended symmetry points. In section \ref{conditionsmodels} we analyse the latter effects and discuss the conditions for constructing semi-realistic models with positive, exponentially suppressed values for the vacuum energy. In section \ref{exmetastable}, we present an example model that satisfies the abovementioned conditions and demonstrate its dynamical stability in a wide region of parameter space. We end in section \ref{conclusions} with our conclusions.


\section{Exploring a class of non-supersymmetric  $(\mathbb{Z}_2)^6$ models}\label{FFFscan}

We have already announced the importance of controlling the behaviour of the one-loop potential at special points in moduli space, where the size of the internal manifold is of the order of the string scale. Our first task is then to examine the possibility of constructing  chiral, tachyon free, non-supersymmetric  heterotic string models with non-negative cosmological constant at such special points of enhanced symmetry. For this purpose, a convenient choice is to utilise the framework of the Free Fermionic Formulation \cite{freeferm} of the heterotic string to scan the space of models possessing various properties of interest. In subsequent sections, we will deform the theory away from the fermionic point, and obtain the full expression for the one-loop effective potential as a function of the no-scale moduli.

A model in the Free Fermionic Formulation is defined by a set of basis vectors  $\{\beta_1,\beta_2,\dots,\beta_N\}$, associated with  the parallel transport properties  of the fermionic coordinates along the two non-contractible loops of the world-sheet torus, and a set of phases $\cc{\beta_i}{\beta_j},$ with $i,j=1,\dots,N$, associated with Generalised GSO (GGSO) projections. We will focus on a particular class of vacua\footnote{In general, due to the presence of radiative corrections to the effective potential which, as we shall see, may lead to runaway solutions, the term `vacuum' is not strictly speaking correct in our non-supersymmetric setup. We shall, however, still employ it by abuse of language.} and take advantage of the formalism developed in \cite{cformalism1,cformalism,cformalism2} in order to derive generic analytic results for the characteristics of these models and subsequently scan for models with the desired properties. We do not provide a comprehensive review of the Free Fermionic Formulation here, but only outline the salient features relevant to our analysis, while referring the reader to \cite{freeferm} and \cite{cformalism1,cformalism,cformalism2} for more details on technical aspects. In later sections, whenever specific models are discussed, their representation as toroidal orbifolds will always be employed.

The  class of vacua under consideration can be described by a fixed set of nine basis vectors $\{\beta_1,\beta_2,\dots,\beta_9\}$ and a variable set of phases  $\cc{\beta_i}{\beta_j}$, where $i,j=1,\dots,9$. Explicitly, one may parametrise the basis vectors in the following convenient form
\begin{align}
	\begin{split}
\beta_1=\textbf{1}&=\{\psi^\mu,\
\chi^{1,\dots,6},y^{1,\dots,6},\omega^{1,\dots,6}|\bar{y}^{1,\dots,6},
\bar{\omega}^{1,\dots,6},\bar{\eta}^{1,2,3},
\bar{\psi}^{1,\dots,5},\bar{\phi}^{1,\dots,8}\}\\
\beta_2=S&=\{\psi^\mu,\chi^{1,\dots,6}\}\\
\beta_3=T_1&=\{y^{12},\omega^{12}|\bar{y}^{12},\bar{\omega}^{12}\}\\
\beta_4=T_2&=\{y^{34},\omega^{34}|\bar{y}^{34},\bar{\omega}^{34}\}\\
\beta_5=T_3&=\{y^{56},\omega^{56}|\bar{y}^{56},\bar{\omega}^{56}\}\\
\beta_6=b_1&=\{\chi^{34},\chi^{56},y^{34},y^{56}|\bar{y}^{34},\bar{y}^{56},
\bar{\psi}^{1,\dots,5},\bar{\eta}^1\}\\
\beta_7=b_2&=\{\chi^{12},\chi^{56},y^{12},y^{56}|\bar{y}^{12},\bar{y}^{56},
\bar{\psi}^{1,\dots,5},\bar{\eta}^2\} \\
\beta_8=z_1&=\{\bar{\phi}^{1,\dots,4}\} \\
\beta_9=z_2&=\{\bar{\phi}^{5,\dots,8}\} ,
	\end{split}\label{basis}
\end{align}
in terms of their explicit worldsheet fermion content. Adopting the conventions of \cite{cformalism1,cformalism,cformalism2}, $\chi^I$ are the six real RNS fermionic superpartners of the internal toroidal directions $X^I$. In the framework of the fermionic construction, the latter are fermionised as $\partial X^I = y^I \omega^I$, and similarly for the right-movers. Moreover, using the fermionic representation of the ${\rm E}_8\times{\rm E_8}$ lattice, the right-moving gauge degrees of freedom are now parametrised by the complex fermions $\bar\psi^{1,\ldots,5}$, $\bar\eta^{1,2,3}$, $\bar\phi^{1,\ldots,4}$, $\bar\phi^{5,\ldots,8}$, associated to ${\rm SO}(10)$, ${\rm U}(1)\times{\rm U}(1)\times{\rm U}(1)$, ${\rm SO}(8)$ and ${\rm SO}(8)$ gauge group factors, respectively.

A set of constraints imposed on the GGSO phases to guarantee modular invariance at one and higher loops, leaves  $2^{9(9-1)/2}+1$ free parameters. As a result, this class comprises $2^{36}+1\sim 10^{11}$  models.   
This formulation has the advantage of leading to explicit expressions in terms of the GGSO coefficients for the basic characteristics of the model, such as the gauge group, the number of fermion families, the cosmological 
constant and the presence of tachyonic states in the case of non-supersymmetric models.
These models enjoy  ${\rm SO}(10)\times {\rm SO}(8)^2\times {\rm U}(1)^9$ gauge symmetry apart from special $\cc{\beta_i}{\beta_j}$ configurations where gauge group enhancements may occur. 
For the purposes of this work we will consider SO(10) as the ``observable" gauge group.
Fermion generations, transforming as SO(10) spinorials, arise from the 
twisted sectors $B^I_{pq}=S+b^I_{pq},$ $I=1,2,3$ where we have introduced  $b^1_{pq}=b^1+p\,T_2+q\,T_3$, $b^2_{pq}=b^2+p\,T_1+q\,T_2$, $b^3_{pq}=x+b^1+b^2+p\,T_1+q\,T_2$, with $p,q\in\{0,1\}$, and $x=\textbf{1}+S+\sum_{i=1}^3 T_i+\sum_{k=1}^2z_k$. The net number of fermion generations is then determined explicitly in terms of the GGSO coefficients to be
\begin{align}
N=\left|\sum_{I=1,2,3} \chi^I\right| \,,
\end{align}
where
\begin{align}
	\begin{split}
\chi^1_{pq}&=-4\,{\rm ch}(\psi^{\mu})\cc{B^1_{pq}}{S+b_2+(1-q)T_3}\,P^1_{pq} \,,\\
\chi^2_{pq}&=-4\,{\rm ch}(\psi^{\mu})\cc{B^2_{pq}}{S+b_1+(1-q)T_3}\,P^2_{pq} \,,\\
\chi^3_{pq}&=-4\,{\rm ch}(\psi^{\mu})\cc{B^3_{pq}}{S+b_1+(1-q)T_1}\,P^3_{pq} \,.
	\end{split}
\end{align}
Here we have introduced the notation ${\rm ch}(\psi^{\mu})$ for the spacetime fermion chirality and  $P^I_{pq}$  for the projectors
\begin{align}
P^I_{pq}&=\frac{1}{2^3}\left(1-\cc{B^I_{pq}}{T_I}\right)\left(1-\cc{B^I_{pq}}{z_1}\right)\left(1-\cc{S+B^1_{pq}}{z_2}\right) \,.
\end{align}

Clearly, the absence of physical (level-matched) tachyons is not guaranteed in the case of non-supersymmetric vacua. Their presence in the string spectrum manifests itself as level-matched singularities in the $q\bar q$-expansion of the one-loop partition function. For a generic fermionic model in the class under consideration, such tachyonic states may either arise from the sectors $z_1,z_2$, in which case they carry conformal weight $\left(-\frac{1}{2},-\frac{1}{2}\right)$, or from the sectors $T_m +p\,z_1+ q\,z_2$, with $m=1,2,3$ and $p,q=0,1$, carrying conformal weights $\left(-\frac{1}{4},-\frac{1}{4}\right)$. Whether such tachyonic states actually appear in the string spectrum of a given model or not, crucially depends on the choice of GGSO projections. Indeed, it turns out that for special choices of GGSO coefficients, the would-be tachyons may be projected out, or acquire mass as a result of the fermionic realisation of the Scherk-Schwarz mechanism, leading to tachyon-free fermionic vacua despite the absence of spacetime supersymmetry.

We note here that this does not imply that the full theory be classically stable. Tachyonic modes may still arise if one deforms the theory away from the fermionic point by marginal operators corresponding to non-trivial (yet constant) VEVs for the metric and $B$-field moduli in the internal toroidal directions. A fully-fledged stability analysis of the various deformations, taking into account the entire parameter space of scalar moduli, is necessary in order to completely identify the tachyonic regions in moduli space. Such a detailed analysis was performed \emph{e.g.} in \cite{Florakis:2010ty}, where specific conditions on the moduli guaranteeing classical stability were obtained for a variety of type II and heterotic theories.

Using the free fermionic formulation, it was possible to obtain general analytic expressions for the partition function expanded in powers of $q,\bar{q}$, valid for all models in the class under consideration, as functions of the GGSO coefficients. Such formulae prove very efficient for computer scans of the space of models, under the condition of absence of tachyonic modes. For instance, the number of tachyonic states with conformal weight $\left(-\frac{1}{2},-\frac{1}{2}\right)$ is given in terms of the GGSO coefficients as
\begin{align}  
 W_{-1/2}=  \frac{1}{16} \left(1+\cc{z_1}{z_2}\right)\times&\sum_{k=1}^2\left[\left(1+\cc{S}{z_k}\right)\prod_{i=1}^3\left(1+\cc{T_i}{z_k}\right) 
 \prod_{a=1}^2\left(1+\cc{b_a}{z_k}\right)\right]\ .
\end{align}
Similar explicit expressions have been obtained for the degeneracy $W_{-1/4}$ of tachyonic modes with conformal weight $\left(-\frac{1}{4},-\frac{1}{4}\right)$. However, they are more 
complicated 
since they receive contributions from several sectors involving the compactification lattice at the fermionic point. As an example, we give here the contribution from the $T_1$ sector
\begin{align}
W_{-1/4}^{T_1}&=\left(1+\cc{S}{T_1}\right)\left[\left(2+\cc{T_1}{z_1}+\cc{S}{z_2}\right)\left(1+\cc{T_1}{b_1}\right)\prod_{i=2}^3\left(1+\cc{T_1}{T_k}\right)\right.
\nonumber\\
&-\frac{3}{2}\left(1+\cc{S}{T_1}\right)\left(1+\cc{T_1}{z_1}\right)\left(1+\cc{T_1}{z_2}\right)\left(1+\cc{T_1}{b_1}\right)\prod_{k=2,3}\left(1+\cc{T_1}{T_k}\right)
\nonumber\\
&+\frac{1}{2}\left(1+\cc{S}{T_1}\right)\left(1+\cc{T_1}{z_1}\right)\left(1+\cc{T_1}{z_2}\right)\left(4+3\cc{T_1}{T_2}+3\cc{T_1}{T_3}+2\cc{T_1}{T_2+T_3}\right) \,.
\end{align}

For our analysis, it is crucial to isolate the $(q,\bar q)$-independent term in the partition function, $W_0\equiv n_B-n_F$, expressed in terms of the GGSO projection coefficients.  This term, ascribed to the difference between massless bosonic and fermionic states,  is directly relevant for determining the sign (and magnitude) of the cosmological constant at the fermionic point. The full expressions for $W_0$ and $W_{-1/4}$ are quite involved and will be reported elsewhere \cite{prep}.

With the help of the analytic expressions for the net chirality $N$, the tachyonic contributions $W_{-1/2},W_{-1/4}$ and the massless supertrace $W_0$, we can launch a computer scan for models with the desired characteristics. It turns out that the number of relevant GGSO coefficients is 27 namely $\cc{S}{T_i},\cc{S}{z_k}$, $\cc{T_i}{T_j},\cc{T_i}{z_k},\cc{b_a}{z_k},\cc{T_1}{b_1},\cc{T_2}{b_2},\cc{T_3}{b_a},\cc{z_1}{z_2}$ and $\cc{\textbf{1}}{S}$,
$\cc{S}{b_a} , \cc{b_1}{b_2}$, with  $a,k=1,2$ and $i,j=1,2,3$. However, the last four are actually redundant, since they are related
to a  flip in the overall chirality sign. Setting  $\cc{S}{T_1}=+1$ in order to restrict our scan to non-supersymmetric vacua, we are left with 22 independent phases.

It is important to note here one additional requirement concerning the specific embedding of the Scherk-Schwarz breaking of supersymmetry into the free-fermionic framework. Our eventual goal is to marginally deform the theory away from the fermionic point at generic VEVs of the compactification moduli which, from the point of view of the fermionic formulation, are kept fixed at special values. We would like the resulting model to have spontaneously broken supersymmetry, with the gravitino mass scale being determined in terms of the volume of the first 2-torus, $m_{3/2}\sim 1/R$. This operation proves to be quite intricate,
 involving  four additional GGSO phases 
and inferring extra constraints on GGSO coefficients.

These 
additional constraints, that guarantee the spontaneous breaking of supersymmetry have been also implemented in the computer-aided search programme. 
The scan eventually involves 26 GGSO phases taking values $\pm1$ and result in a total of $2^{26}\sim10^{8}$ models. Although a comprehensive scan is possible, for
the purposes of this work, we instead chose to perform a random scan in a sample of $10^6$ models. 
We search for models that meet the following criteria:
\begin{itemize}
 	\item Absence of physical tachyons, $W_{-1/2}=W_{-1/4}=0$.
	\item Presence of chiral matter,  $N\ne0$.
	\item Spontaneous SUSY breaking consistent with the Scherk-Schwarz mechanism of field theory, and with the breaking scale controlled by the volume of the first 2-torus.
\end{itemize}
Moreover, using the techniques presented in section \ref{examplemodel}, we have computed
the difference between massless bosonic and fermionic states at the generic point ($W_{0}^G$) in order to enumerate super no-scale models in this class. 

\begin{table*}\centering
\ra{1.3}
\begin{tabular}{@{}rrrr@{}}\toprule
 & \# models in the random scan  && estimated \#  models in this class \\ \midrule
$W_0>0, W_{0}^G\ne0$ &  958 && $6.4\times10^4$\\
$W_0>0, W_{0}^G=0$   &  115 && $7.7\times10^3$\\
$W_0<0$             &   62 && $4.2\times10^3$\\ \midrule
Total \# of models&1135 && $7.6\times10^4$\\
\bottomrule
\end{tabular}
\caption{{\it Results of a random scan for models that satisfy all constraints over a sample of $10^6$ models in the class under consideration (comprising $6.7\times10^7$ models).} }\label{tablei}
\end{table*}

\begin{figure}
\includegraphics[width=\textwidth]{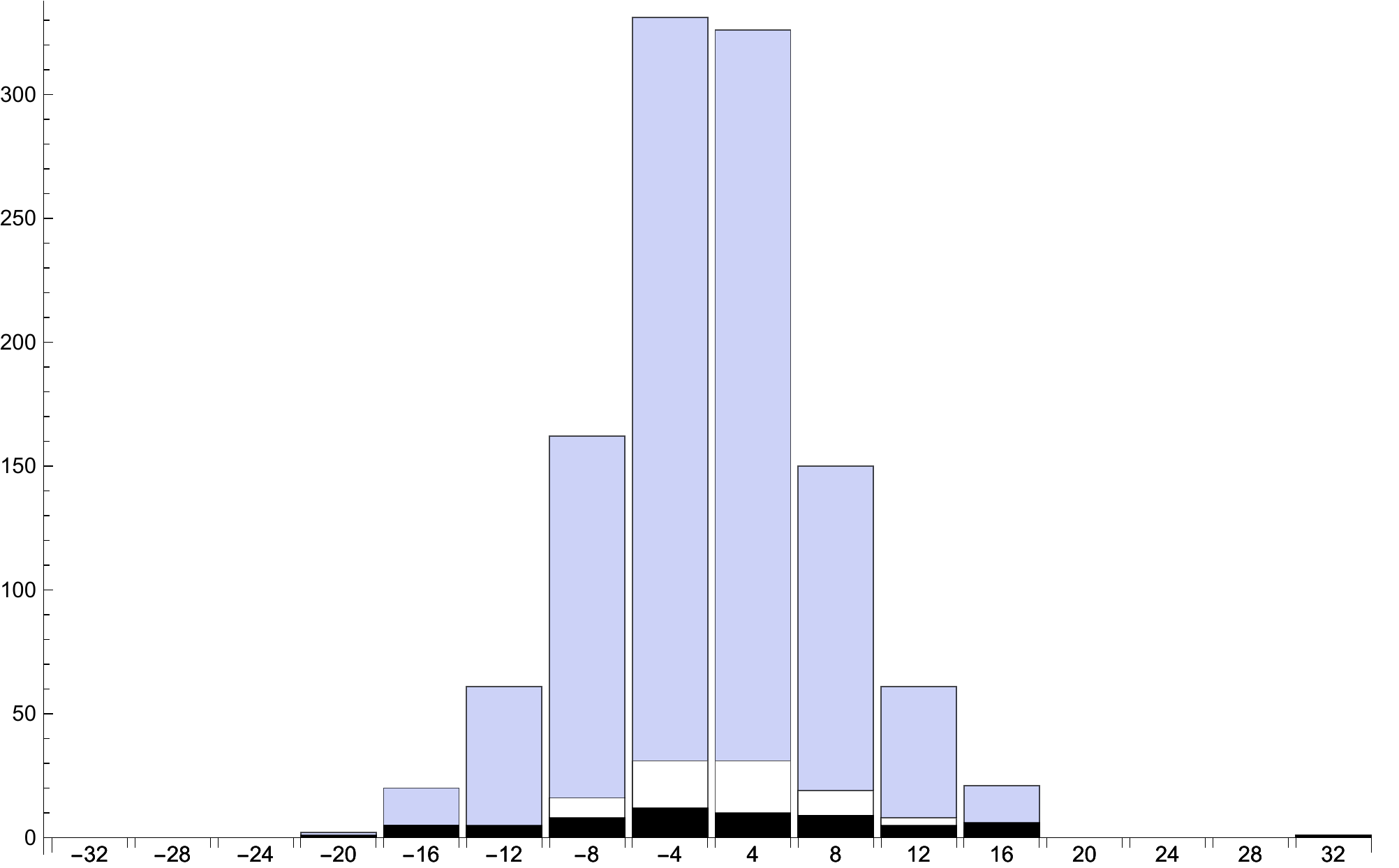}
\caption{\label{fig}{\it Number of acceptable models in the class under consideration versus the net number of fermion families. Black bars represent models with $W_0<0$, while white bars depict super no-scale models ($W_0=0$ at the generic point). The light-coloured bars depict the total number of models satisfying all constraints for every value of the net chirality.}}
\end{figure}

We obtained 1135 models that satisfy all criteria and our results are summarised in Table
 \ref{tablei}. In  Figure \ref{fig}, we plot  the number of models versus the number of fermion generations. The light-shaded columns  correspond to the total number of models satisfying all constraints for each value of net chirality $N$. Black bars  depict models with $W_0<0$ and white columns correspond to super no-scale models $W_0^G=0$, in that, upon deformation away from the fermionic point, they exhibit a Bose-Fermi degeneracy in the massless sector. Our random scan did not produce any models with $W_0=0$ at the fermionic point.

The statistical dominance of models with $W_0>0$, covering 95\% of the models plotted, is striking. As outlined in the introduction, these cases naively are expected to correspond to a negative cosmological constant, with the one-loop potential having the form of a ``puddle" which stabilises the no-scale moduli at the string scale. It turns out that this reasoning, motivated from field theory, is not always correct as we shall see in later sections. On the other hand, models with $W_0<0$, suggesting a positive value for the cosmological constant, only arise in 5\% of the plotted results. 


\section{Model A : an example with positive cosmological constant}\label{examplemodel}

\subsection{Definition and partition function}

In this section we present and analyse in detail one of the models constructed in the previous section, subject to the condition  $n_F>n_B$ at the fermionic point. Since all models under consideration can be analysed in a similar way, this section will also serve to set the notation  and describe most of the general characteristics satisfied also by the models we construct in subsequent sections. The model in question shall be henceforth referred to as `Model A'. It has net chirality $N=12$ and is defined by the GGSO matrix
\begin{align}
	c_{\rm(A)}[^{\beta_i}_{\beta_j}] = \begin{pmatrix}
 1 & 1 & 1 & 1 & -1 & 1 & 1 & 1 & 1 \\
 1 & 1 & 1 & -1 & -1 & 1 & 1 & -1 & 1 \\
 1 & 1 & -1 & 1 & 1 & -1 & 1 & 1 & -1 \\
 1 & -1 & 1 & -1 & -1 & 1 & -1 & -1 & -1 \\
 -1 & -1 & 1 & -1 & 1 & 1 & 1 & -1 & -1 \\
 1 & -1 & -1 & 1 & 1 & 1 & 1 & -1 & 1 \\
 1 & -1 & 1 & -1 & 1 & 1 & 1 & -1 & -1 \\
 1 & -1 & 1 & -1 & -1 & -1 & -1 & 1 & -1 \\
 1 & 1 & -1 & -1 & -1 & 1 & -1 & -1 & 1\\
			\end{pmatrix} \,.
			\label{firstexampleGSO}
\end{align}
Being one of the solutions to our computer-aided scan, it satisfies all the conditions and requirements introduced in the previous section. In particular, at the fermionic point, it has no physical tachyons and its massless spectrum is dominated by fermions. Indeed, the expansion of its partition function in powers of $q,\bar q$ at the fermionic point displays these features
\begin{align}
	Z=-312+\frac{2}{\bar{q}}+\frac{56 q}{\bar q}-\frac{16 q^{1/4}}{\bar{q}^{3/4}}+\frac{32 \sqrt{q}}{\sqrt{\bar{q}}}+\frac{512 q^{3/4}}{\bar{q}^{1/4}}+4064 q^{1/4} \bar{q}^{1/4}+12288 \sqrt{q} \sqrt{\bar{q}}+\frac{6144 \bar{q}^{3/4}}{q^{1/4}}+\ldots
		\label{expansionfff}
\end{align}
The important observation here is the negative sign of the constant term, indicating an abundance of fermions at the massless level. Note also the absence of level-matched physical tachyons of the form $(q\bar q)^{-1/2}$ or $(q\bar q)^{-1/4}$. 

In order to study the structure of the one-loop effective potential, we need to perturb the theory away from the fermionic point by marginally deforming the sigma model with current-current operators that re-introduce the dependence on the compactification moduli. This can be most conveniently achieved by re-writing the model as an orbifold theory. We define generically the K\"ahler and complex structure moduli of each 2-torus as $T^{(i)}=T^{(i)}_1+iT^{(i)}_2$ , $U^{(i)}=U^{(i)}_1+i U^{(i)}_2$.

The one-loop partition function of the model at the generic point in the perturbative heterotic moduli space reads
\begin{align}
	\begin{split}
	Z = &\frac{1}{\eta^{12}\bar\eta^{24}}\,\frac{1}{2^3}\sum_{h_1,h_2,H\atop g_1,g_2,G}\frac{1}{2^3}\sum_{a,k,\rho\atop b,\ell,\sigma} \frac{1}{2^3}\sum_{H_1,H_2,H_3\atop G_1,G_2,G_3} (-1)^{a+b+HG+\Phi}\, \vartheta[^a_b]\,\vartheta[^{a+h_1}_{b+g_1}]\,\vartheta[^{a+h_2}_{b+g_2}]\,\vartheta[^{a-h_1-h_2}_{b-g_1-g_2}] \\
			&  \times  \Gamma^{(1)}_{2,2}[^{H_1}_{G_1}|^{h_1}_{g_1}](T^{(1)},U^{(1)})\ \Gamma^{(2)}_{2,2}[^{H_2}_{G_2}|^{h_2}_{g_2}](T^{(2)},U^{(2)})\ \Gamma^{(3)}_{2,2}[^{H_3}_{G_3}|^{h_1+h_2}_{g_1+g_2}](T^{(3)},U^{(3)}) \\
		& \times	\ \bar\vartheta[^k_\ell]^5 \,\bar\vartheta[^{k+h_1}_{\ell+g_1}]\,\bar\vartheta[^{k+h_2}_{\ell+g_2}]\,\bar\vartheta[^{k-h_1-h_2}_{\ell-g_1-g_2}]
			 \,\bar\vartheta[^{\rho}_{\sigma}]^4 \,\bar\vartheta[^{\rho+H}_{\sigma+G}]^4 \,,
	\end{split}
	\label{orbifoldpartitionf}
\end{align}
where $\eta(\tau)$ is the Dedekind eta function, $\vartheta[^\alpha_\beta](\tau)$ is the Jacobi theta constant with characteristics, and $a,b$ are the spin structures associated to the worldsheet fermions of the orbifold theory, with the NS sector corresponding to $a=0$ and the R sector to $a=1$. The parameters $k,\ell=0,1$ and $\rho,\sigma=0,1$ label the boundary conditions of the 16 complex fermions realising the level one Kac-Moody algebra of each ${\rm E}_8$ factor of the ten-dimensional heterotic string, respectively. Namely, the fermionic realisation of the ${\rm E}_8\times{\rm E}_8$ lattice in this notation reads
\begin{align}
	\Gamma_{{\rm E}_8\times{\rm E}_8} = \frac{1}{2}\sum_{k,\ell=0,1} \bar\vartheta[^k_\ell]^8 \ \frac{1}{2}\sum_{\rho,\sigma=0,1} \bar\vartheta[^{\rho}_{\sigma}]^8 \,,
\end{align}
and we adopt the convention that the left movers, associated to the holomorphic $\tau$ parameter,  denote the RNS side of the heterotic string. 

Furthermore, $(h_1, g_1)$ and $(h_2,g_2)$ are the orbifold parameters for the $\mathbb Z_2\times\mathbb Z_2$ non-freely acting orbifold with standard embedding generating a chiral $\mathcal N=1$ theory which may be thought of as the singular limit of a Calabi-Yau manifold. In this language, the  $h_i=0,1$ label the various orbifold (un)twisted sectors, while summation over $g_i$ imposes the associated invariance projections. Similarly, $H_i, G_i=0,1$ are generically associated to three freely-acting  $\mathbb Z_2$  orbifolds, each involving an order-two  shift on one of the three $T^2$ tori, and are associated with the Scherk-Schwarz breaking. Finally, $H,G=0,1$ are associated to the orbifold twisting 4 Kac-Moody currents in the hidden ${\rm E}_8$ directions and breaking it down to ${\rm SO}(8)\times {\rm SO}(8)$.

The phase $\Phi$ corresponding to the specific choice $c[^{\beta_i}_{\beta_j}]$ of the fermionic construction is given by
\begin{align}
	\begin{split}
	\Phi =& ab + k\ell + \rho \sigma\\
		   & + ag_2+b h_2+h_2 g_2\\
		   & + k G + \ell H+ HG\\
		   & + k G_2 + \ell H_2 + H_2 G_2\\
		   & +G_1(a+\rho)+H_1(b+\sigma)\\
		   & +(G_1+G_2)H + (H_1+H_2)G\\
		   & +G_3 H_2 + H_3 G_2\\
		   & + h_1 g_2+ g_1 h_2\\
		   & + h_1 G + g_1 H\\
		   & + G_1 h_2 + H_1 g_2\\
		   & + G_2 (h_1+h_2)+H_2(g_1+g_2)\\
		   & +G_3 h_1 + H_3 g_1 \,.
	\end{split}
\end{align}
It is invariant under modular transformations and its exact determination can be straightforwardly obtained using the techniques outlined in \cite{IoannisFlorakis:2011wfd}. This phase implements the Scherk-Schwarz breaking and the consistent embedding of the action of the GGSO projections of the fermionic construction into the Narain lattices and sectors of the orbifold theory.

Equivalently, the model can be constructed as the orbifold compactification on $T^6 / (\mathbb{Z}_2)^6$, where we associated
 each of the six $\mathbb Z_2$ factors with their corresponding parameters
\begin{align}
	\begin{split}
		& \mathbb Z_2^{(1)} \ : \ (h_1,g_1)\\
		& \mathbb Z_2^{(2)} \ : \ (h_2,g_2)\\
		& \mathbb Z_2^{(3)} \ : \ (H_1,G_1)\\
		& \mathbb Z_2^{(4)} \ : \ (H_2,G_2)\\
		& \mathbb Z_2^{(5)} \ : \ (H_3,G_3)\\
		& \mathbb Z_2^{(6)} \ : \ (H,G) \,.
	\end{split}
	\label{z6param}
\end{align}
Their action on the worldsheet degrees of freedom can be summarised\footnote{Although the model can be equivalently described in simpler terms as an orbifold with fewer $\mathbb Z_2$ factors, we prefer to adopt the current presentation, since it generalises most straightforwardly to describe also the models that we discuss in later sections.} as  follows:
\begin{align}
	\begin{split}
		& \mathbb Z_2^{(1)} \ : \ X^{1,2,5,6}\to - X^{1,2,5,6}\\
		& \mathbb Z_2^{(2)} \ : \ X^{3,4,5,6}\to - X^{3,4,5,6}\\
		& \mathbb Z_2^{(3)} \ : \ (-1)^{F_{\rm s.t.}+F_2}\,\delta_1 \quad,\ \delta_1: \{ X_1\to X_1+\pi R_1\} \\
		& \mathbb Z_2^{(4)} \ : \ (-1)^{F_1}\,\delta_3 \quad,\ \delta_3: \{ X_3\to X_3+\pi R_3\} \\
		& \mathbb Z_2^{(5)} \ : \ \delta_5 \quad,\ \delta_5:\{ X^5\to X^5+\pi R_5\} \\
		& \mathbb Z_2^{(6)} \ : \ (-1)^{F_1}\,r \quad,\ r:\{\bar\phi^{5,6,7,8}\to -\bar\phi^{5,6,7,8}\} \,,
	\end{split}\label{modelAorbdef}
\end{align}
where $F_{\rm s.t.}$ is the spacetime fermion number and $F_1, F_2$ are the `fermion' numbers associated to the spinorial representations of the two ${\rm E}_8$ gauge group factors, respectively. Finally, matching with the fermionic construction of the model further requires turning on non-trivial discrete torsion as follows
\begin{align}
	\epsilon(1,2),\ \epsilon(1,4),\ \epsilon(1,5),\ \epsilon(1,6),\ \epsilon(2,3),\ \epsilon(2,4),\ \epsilon(3,6),\ \epsilon(4,5),\ \epsilon(4,6) \,,
\end{align}
where $\epsilon(i,j)$ denotes the discrete torsion assignment between the orbifolds $\mathbb{Z}_2^{(i)}$ and $\mathbb Z_2^{(j)}$.

Special care is needed when treating the contribution of the $(2,2)$ toroidal lattices. The combined orbifold action involves shifts and twists along the same directions of the internal 2-tori, which may cause certain sectors to vanish identically. For example, if the lattice is twisted with respect to a shift orbifold, it produces states with non-trivial momenta and windings. However, introducing also a non-trivial element of a rotation orbifold inside the trace yields a vanishing result since it projects onto states with vanishing momenta and windings. At the level of the partition function, the twisted/shifted (2,2) lattices  $\Gamma_{2,2}[^{H_i}_{G_i}|^{h}_{g}](T,U)$ are then given by
\begin{align}
	\Gamma_{2,2}[^{H_i}_{G_i}|^{h}_{g}]	(T,U) = \begin{cases}
						\bigr|\frac{2\eta^3}{\vartheta[^{1-h}_{1-g}]}\bigr|^2 & , \  (H_i,G_i)=(0,0)\ {\rm or}\ (H_i,G_i)=(h,g) \\
						\Gamma_{2,2}^{\rm shift}[^{H_i}_{G_i}](T,U) & , \  h=g=0  \\
						0 & , \ {\rm otherwise}\\
	\end{cases}\,,
\end{align}
and the partition function of the (2,2) shifted Narain lattice itself is defined\footnote{This particular action on the lattice, in terms of a momentum shift along the torus $a$-cycle together with a winding shift along the $b$-cycle might appear to be different from the simple definition \eqref{modelAorbdef} in terms of a single momentum shift along the $a$-cycle. The two are, however, related by a redefinition $T\to T-1$, as we discuss later in this section.} as
\begin{align}
	\Gamma_{2,2}^{\rm shift}[^{H_i}_{G_i}](T,U) = \sum_{m_1,m_2\atop n_1,n_2} (-1)^{G(m_1+n_2)}\, q^{\frac{1}{4}|P_L|^2}\,\bar{q}^{\frac{1}{4}|P_R|^2} \,,
\end{align}
with momentum shift along the 1rst cycle and winding shift along the 2nd cycle (case VIII with $\lambda=0$ in the classification of \cite{Gregori:1997hi}), so that the lattice momenta are given by
\begin{align}
	\begin{split}
	& P_L= \frac{m_2+\frac{H_i}{2} - U m_1 + T (n_1+\frac{H_i}{2}+U n_2)}{\sqrt{T_2 U_2}}  \,, \\
	& P_R= \frac{m_2+\frac{H_i}{2} - U m_1 +\bar T (n_1+\frac{H_i}{2}+U n_2)}{\sqrt{T_2 U_2}} \,. 
	\end{split}
\end{align}
Matching the orbifold partition function with the one obtained by means of the fermionic construction may be achieved straightforwardly by noticing that the fermionic point corresponds to setting the toroidal moduli to the values $T=i$ and $U=(1+i)/2$, at which point the twisted/shifted (2,2) lattice can be represented entirely in terms of theta functions
\begin{align}
	\Gamma_{2,2}[^{H_i}_{G_i}|^{h}_{g}]	(T=i,U=\tfrac{1+i}{2}) = \frac{1}{2}\sum_{\epsilon,\zeta} \Bigr|\vartheta[^{\epsilon}_{\zeta}]\,\vartheta[^{\epsilon+h}_{\zeta+g}]\Bigr|^2\ (-1)^{H_i(\zeta+g)+G_i(\epsilon+h)+H_i G_i} \,.
	\label{fermlattice}
\end{align}
It is a non-trivial check that the orbifold partition function  eq. (\ref{orbifoldpartitionf}) does indeed reproduce the same $q,\bar q$ expansion as the one produced by the fermionic construction, eq. (\ref{expansionfff}) at this special point in moduli space.

\subsection{Gravitino mass}

Let us now discuss the gravitino mass of the model. Clearly, this originates from the sector
\begin{align}
	k=\rho=H=h_1=h_2=H_1=H_2=H_3=0 \quad ,\quad a=1 \,,
\end{align}
in which the phase simplifies to $\Phi=b+g_2+G_1$. The projections over $b$, $g_1$ and $g_2$ simply pick the $\mathcal N=1$ gravitino and fix its chirality, and similarly for other projections. The non-trivial part is the summation over $G_1$ which imposes $m_1+n_2$ to be odd. Given that we must have $|P_L|=|P_R|$ for level matching, we must also set $n_1=n_2=0$ and, therefore $m_1$ is now constrained to be odd. The lowest mass state is therefore associated to the gravitino, and corresponds to taking $m_1=\pm 1$ and $m_2=0$, giving
\begin{align}
	m_{3/2} = \frac{|U^{(1)}|}{\sqrt{T_2^{(1)} U_2^{(1)}}} \,,
	\label{gravitinomass}
\end{align}
which is the standard Scherk-Schwarz mass term one would have expected from field theory, \emph{i.e.} for a square torus $T=iR_1 R_2$, $U=iR_2/R_1$ one finds $m_{3/2}=1/R_1$. At the fermionic point,  setting $T^{(1)}=i$ and $U^{(1)}=(1+i)/2$, one finds $m_{3/2}=1$ in string units. This is again consistent with the analogous result from the fermionic construction, which implies that the gravitino comes from the term $q^{1/4}\bar q^{1/4}$ with conformal weights $(\frac{1}{4},\frac{1}{4})$.

\subsection{Deformation directions and  T-duality}

As explained in previous sections, our main goal is to study the behaviour of the one-loop effective potential for the model in question. While the framework of the fermionic construction is very helpful for constructing models with certain desired phenomenological properties such as chirality, it only lives at a special point in moduli space and is clearly not suited for the study of the shape of the potential as a function of the moduli. To this end, the previous subsection was devoted to the  rewriting of the model as an orbifold. 

Before directly investigating the actual shape of the one-loop potential, it is important to identify the directions in moduli space which are relevant for our analysis. As outlined in the Introduction, these are primarily the toroidal moduli in the directions controlling the Scherk-Schwarz breaking of supersymmetry. In all models, we conventionally choose them to  be the K\"ahler and complex structure moduli $T^{(1)}$ and $U^{(1)}$ associated to the first 2-torus.

Without the shifts, each 2-torus enjoys the full ${\rm SL}(2;\mathbb Z)_T \times {\rm SL}(2;\mathbb Z)_U \ltimes \mathbb Z_2$ T-duality symmetry. The action of the shifts, however, breaks this into a subgroup which typically depends on the specific way the orbifold shifts the left- and right- moving toroidal coordinates. Moreover, since fixed points under T-duality correspond to extrema of the one-loop potential,  it is important to discuss the residual T-duality group as this will play an important role for our later analysis of the one-loop potential.

Notice first, that an ${\rm SL}(2;\mathbb Z)_T$ translation may be employed in order to rewrite the shifted lattice $\Gamma$ as
\begin{align}
	\Gamma^{\rm shift}_{2,2}[^{H_i}_{G_i}](T,U)= {\Gamma^{\rm shift}_{2,2}}'[^{H_i}_{G_i}](T-1,U)\,,
\end{align}
where $\Gamma'$ is a new shifted lattice with a single momentum shift along the first cycle. The T-duality group for each 2-torus is identified to be $	\Gamma^1(2)_T \times \Gamma_0(2)_U$, with  $\Gamma_0(2)$ being the  Hecke congruence subgroup of ${\rm SL}(2;\mathbb Z)$, 
\begin{align}
	\Gamma_0(2) = \left\{ \begin{pmatrix}
					a & b\\
					c & d\\
				\end{pmatrix} \in {\rm SL}(2;\mathbb Z)\ |\ c=0\,{\rm mod}\,2\right\} \,,
\end{align}
and
\begin{align}
	\Gamma^1(2) = \left\{ \begin{pmatrix}
					a & b\\
					c & d\\
				\end{pmatrix} \in {\rm SL}(2;\mathbb Z)\ |\ a,d=1\,{\rm mod}\,2\ \ {\rm and}\ \ b=0\,{\rm mod}\, 2\right\} \,,
\end{align}
where group elements $\gamma=\binom{ a\ b}{c\ d}$ act on the modulus $z$ via fractional linear transformations $z \to (a z+b)/(cz+d)$.

An inspection of the gravitino mass \eqref{gravitinomass} reveals that the relevant parameter controlling the SUSY breaking scale is the volume of the first 2-torus, $T_2^{(1)} \equiv {\rm Im}(T^{(1)})$. For simplicity, we will therefore deform the theory away from the fermionic point ($T^{(1)}_2=1$) in this direction, while keeping all other moduli frozen at their fermionic values 
\begin{align}
	T^{(1)} = -1 + i T_2 \quad,\quad T^{(2)}=T^{(3)}=-1+i \quad,\quad U^{(i)}=\frac{1+i}{2}\,.
	\label{modulideformation}
\end{align}
We henceforth focus on the first 2-torus and, unless otherwise stated,  we suppress the explicit label $T^{(1)}\to T$.  It is  then easy to see that the  $\Gamma^1(2)_T$ transformation
\begin{align}
	\begin{pmatrix}
					-1 & -2 \\
					1 & 1 \\
	\end{pmatrix} \in \Gamma^1(2)_T \quad \leftrightarrow\quad  T \to -\frac{T+2}{T+1} \,,
\end{align}
is a symmetry of the lattice and  transforms the $T$ variable as
\begin{align}
	-1+i T_2 \to -1+ i \frac{1}{T_2} \,.
\end{align}
This means that, along our chosen deformation direction, the potential will exhibit a T-duality symmetry $T_2 \to 1/T_2$, with the fermionic point $T_2=1$ indeed corresponding to the fixed point. We will see in the following that, as expected, the fermionic point corresponds to a maximum of the potential, leading to a dynamical roll to the regime $T_2\gg 1$.

\subsection{Evaluation of the one-loop potential and asymptotics}

We are now ready to study the actual form of the one-loop potential as a function of $T_2$, as outlined above. A complete analysis of the potential as a function of all 6 complex $T^2$ moduli, together with all allowed Wilson line deformations is a daunting task that lies outside the scope of this work. Instead, we will concentrate on its functional dependence on the volume modulus $T_2$ controlling the supersymmetry breaking. 

Even in this simplified setup, however, the analysis of the potential around the fermionic point is still quite involved. As discussed in the Introduction, the potential is obtained as an integral of the partition function of the theory \eqref{potentialdef} over the fundamental domain
\begin{align}
	\mathcal F = \{ \tau \in \mathbb{C}^{+} \ {\rm with}\ |\tau|^2 >1\ {\rm and}\ |\tau_1|\leq \tfrac{1}{2} \} \,.
\end{align}

Much of the difficulty in evaluating such modular integrals stems from the non-rectangular shape of $\mathcal F$, and involves  also the contribution of non-level matched states, as required by unitarity and modular invariance. The traditional method for evaluating this integrand is known in the physics literature as unfolding \cite{McClain:1986id,O'Brien:1987pn,Dixon:1990pc}, and relies on decomposing the Narain lattice contained in $Z(\tau,\bar\tau)$ into orbits under the modular group. The integral in each orbit may then be traded for an integral over a single coset representative. The sum of modular transformations acting on a single representative within each orbit produces a union of images of $\mathcal F$ under elements $\gamma$ of the modular group,  reconstructing a rectangular integration region which is typically easier to evaluate.

Unfortunately, for non-BPS amplitudes such as the effective potential, aside from the Narain lattice, the integrand involves the non-holomorphic dependence on the modular parameter arising from the infinite tower of string oscillators, $ \eta^{-12}(\tau) \bar\eta^{-24}(\bar\tau) $. The degeneracy of the latter grows exponentially as $\sim e^{c\sqrt{n}}$ with $n$ being the mass level and $c$ being a positive constant, and poses convergence issues for values of the torus volume $T_2$ sufficiently close to the string scale. An excellent discussion of these issues can be found in \cite{Angelantonj:2010ic}.

In what follows, we will extract the asymptotic behaviour of the effective potential by direct unfolding of the Narain lattice and contrast it with a numerical evaluation of the potential. A considerable simplification of the expression for the partition function may be obtained by rewriting the lattices associated to the second and third 2-torus entirely in terms of theta functions and by performing the sum over spin structures $(a,b)$ encoding the R-symmetry charges using the Riemann-Jacobi identity
\begin{align}
	\frac{1}{2}\sum_{a,b} (-1)^{a(1+G_1)+b(1+H_1)}\,\vartheta[^a_b]\,\vartheta[^{a+h_1}_{b+g_1}]\,\vartheta[^{a+h_2}_{b+g_2}]\,\vartheta[^{a-h_1-h_2}_{b-g_1-g_2}]=\vartheta[^{1+H_1}_{1+G_1}]\, \vartheta[^{1+H_1+h_1}_{1+G_1+g_1}]\, \vartheta[^{1+H_1+h_2}_{1+G_1+g_2}]\, \vartheta[^{1+H_1-h_1-h_2}_{1+G_1-g_1-g_2}]\,.
	\label{Riemannident}
\end{align}
An additional simplification arises if one notices  that only the sector $h_1=g_1=0$ yields a non-vanishing contribution to the partition function.
Clearly, if $(H_1,G_1)=(0,0)$ the partition function vanishes, because $\mathcal N=1$ supersymmetry is recovered. Therefore, non-trivial contributions only arise from $(H_1,G_1)\neq(0,0)$. However, if $(H_1,G_1)=(h_1,g_1)$ then again, supersymmetry is effectively recovered because the model becomes essentially equivalent to a spontaneous breaking of $\mathcal N=2\to 1$, as can be inferred from \eqref{Riemannident}. Therefore, $(H_1,G_1)\neq (h_1,g_1)$ as well. However, the twisted/shifted lattice $\Gamma_{2,2}^{(1)}[^{H_1}_{G_1}|^{h_1}_{g_1}]$ vanishes identically due to the combined twist and shift, unless $h_1=g_1=0$. We emphasise here that this simplification is a numerical one, and is useful only for the evaluation of the vacuum energy. Computations of interaction terms, such as the running of gauge couplings,  instead see the full structure of the partition function and receive non-trivial contributions also from the sectors $(h_1,g_1)\neq (0,0)$.

Taking these simplifications into account, we extract the asymptotic behaviour at large volume as follows. We first set $h_1=g_1=0$, replace the lattices of the second and third 2-torus in terms of theta functions using \eqref{fermlattice} and finally perform the sum over the $(a,b)$ spin structures. Secondly, we focus on the generic sector $(H_1,G_1)$ and define the orbifold block of the lattice-independent coefficient
\begin{align}
	\begin{split}
	\Psi[^{H_1}_{G_1}] = & \frac{1}{\eta^{12} \,\bar\eta^{24}} \frac{1}{2^8} \sum_{h_2,H =0,1 \atop g_2,G=0,1}\sum_{k,\rho,\gamma_2,\gamma_3=0,1 \atop \ell,\sigma,\delta_3,\delta_3=0,1} (-1)^{ \hat\Phi} \\
	      & \times \vartheta[^{1+H_1+h_2}_{1+G_1+g_2}]^2\,\vartheta[^{1+H_1}_{1+G_1}]^2\\
	      & \times \bar\vartheta[^k_\ell]^6 \, \bar\vartheta[^{k+h_2}_{\ell+g_2}]^2\,\bar\vartheta[^{\rho}_{\sigma}]^4 \, \bar\vartheta[^{\rho+H}_{\sigma+G}]^4 \\
	      & \times \vartheta[^{\gamma_2}_{\delta_2}] \,\vartheta[^{\gamma_2+h_2}_{\delta_2+g_2}]\,\bar\vartheta[^{\gamma_2}_{\delta_2}] \,\bar\vartheta[^{\gamma_2+h_2}_{\delta_2+g_2}]\\
	      &\times \vartheta[^{\gamma_3}_{\delta_3}]\,\vartheta[^{\gamma_3-h_2}_{\delta_3-g_2}]\,\bar\vartheta[^{\gamma_3}_{\delta_3}]\,\bar\vartheta[^{\gamma_3-h_2}_{\delta_3-g_2}] 
	\end{split}
	\label{Zsimplified}
\end{align}
where the phase $\hat \Phi$ is given by
\begin{align}
	\begin{split}
	\hat\Phi = &  (\rho +H+h_2)(\sigma +G+g_2)+G_1(\rho +H+h_2)+H_1(\sigma +G+g_2)\\
	&+\gamma_2(\ell+\delta_3+G)+\delta_2(k+\gamma_3+H)+\gamma_2 \delta_2\\
	&+\gamma_3(\ell+G)+\delta_ 3(k+H)+H(\sigma +G)+G(\rho +H)\\
	&+ (\gamma_2+\gamma_3+k+\rho )g_2+(\delta_ 2+\delta_3+\ell+\sigma )h_2+h_2 g_2\\
	& + g_2(1+H_1+k+h_2) \,.
	\end{split}
\end{align}
Since we fixed the moduli of the second and third torus to their fermionic point values, $\Psi$  is now moduli independent and  is written entirely in terms of theta  and Dedekind functions. It will be convenient to expand it in powers of the real and imaginary parts 
\begin{align}
	q_{\rm r} = e^{-2\pi \tau_2} \quad,\quad q_{\rm i} = e^{2\pi i \tau_1} \,,
\end{align}
as follows
\begin{align}
	\Psi[^{H_1}_{G_1}] = \sum_{M,N} c[^{H_1}_{G_1}](N,M)\,\,q_{\rm r}^N \, q_{\rm i}^M \,,
\end{align}
in terms of  the expansion coefficients $c[^{H_1}_{G_1}](N,M)$ in each orbifold sector $(H_1,G_1)$.

The one-loop effective potential may then be obtained by exactly evaluating the integral \eqref{potentialdef} using the unfolding method. This requires one to make a choice of Weyl chamber, in order to ensure the absolute convergence necessary for exchanging the integral with the sum over the various modular transformations that unfold $\mathcal F$ into a rectangular region. For our purposes, the natural choice is $T_2>1$. The result of the integral in the chamber $T_2<1$ can  be similarly obtained by first Poisson-resumming the lattice to go to the dual frame $T_2'=1/T_2$, and then unfolding again for $T_2'>1$. 

The issue of absolute convergence around $T_2=1$ forcing us to pick a particular chamber in order to justify the unfolding against the Narain lattice is one of the drawbacks of this method and the resulting expression it produces is no longer manifestly invariant under T-duality. In particular, this signals the inability of this unfolding method to capture the behaviour around self-dual points and should be thought of only as an asymptotic expansion valid at sufficiently large volume. 

With this in mind, we proceed with the unfolding of the integral, and the result is broken into the contributions of the various orbits
\begin{align}
	-2(2\pi)^4\,V_{\rm one-loop}(T,U) = I[^0_0]+ I_{\rm deg}[^0_1] + I_{\rm nd}[^0_1] + I_{\rm nd}[^1_0] + I_{\rm nd}[^1_1] \,.
\end{align}
The orbit $I[^0_0]$ vanishes identically due to the spontaneous nature of the breaking of supersymmetry and one is left to evaluate the degenerate orbit $I_{\rm deg}[^0_1]$ and the non-degenerate orbit $I_{\rm nd}[^0_1]+I_{\rm nd}[^1_0]+I_{\rm nd}[^1_1]$.

The $I_{\rm nd}$ contributions are exponentially suppressed in $T_2$, at least by a factor $e^{- 2\pi T_2}$ and, hence, can be safely disregarded away from $T_2=1$. The asymptotic behaviour of the potential is therefore dominated by the contribution $I_{\rm deg}[^0_1]$ of the degenerate orbit
\begin{align}
	\begin{split}
	I_{\rm deg}[^0_1] &= \frac{2c[^0_1](0,0)}{\pi^3 T_2^2} \sum_{m_1,m_2\in\mathbb Z} \frac{U_2^3}{\left|m_1+\frac{1}{2}+U m_2\right|^6}\\
	&+ \frac{4\sqrt{2}}{\sqrt{T_2}}\sum_{N\geq 1} N^{3/2} \,c[^0_1](N,0)\, \sum_{m_1,m_2\in\mathbb Z}\frac{U_2^{3/2}}{ \left|m_1+\frac{1}{2}+U m_2 \right|^3 }\,K_3\left(2\pi \sqrt{ \frac{N T_2}{U_2} \left|m_1+\frac{1}{2}+U m_2\right|^2} \right) \,,
	\end{split}
	\label{AsymptoticExpr}
\end{align}
where $K_s(z)$ is the modified Bessel function of the second kind. 

Let us now discuss the structure of this result. The contribution in the first line is due to states that remain massless at the generic point in the $(T,U)$ moduli space and exhibits the power-like suppression $1/T_2^2$. This falls asymptotically like $\sim 1/R_1^4$, with $R_1$ being the radius of the Scherk-Schwarz circle, as expected from dimensional arguments\footnote{Generically, the dominant behaviour $1/R^4$  persists to all orders in perturbation theory, \emph{c.f.} \cite{Antoniadis:1998sd}.}. From the point of view of a Euclidean treatment, the identification of the compactified time direction with the circle of radius $R_1$ would give rise to a thermal deformation of the theory, with the one-loop potential being mapped to the free energy of the system in a thermal bath of temperature $1/R_1$. The $1/R_1^4$ behaviour would then be interpreted as the Stefan-Boltzmann law.

The shape of the dominant asymptotic suppression is precisely controlled by the difference  $c[^0_1](0,0)=n_B-n_F$ between the number of bosons and fermions that remain massless at the generic point in the $(T,U)$ moduli space. For the model under consideration, it is found to be $c[^0_1](0,0)=-256$. This abundance of fermions implies that the asymptotic suppression of the potential $V_{\rm one-loop}$ at large $T_2\gg 1$ is given by a convex, positive function of the volume. We then expect this to be continuously and smoothly connected to the maximum of the curve, achieved at the fermionic point $T_2=1$, although the latter is not captured by the above asymptotic expression. 

On the other hand, an abundance of massless bosons would have instead implied that the asymptotics be dominated by a concave negative function of $T_2$, and would have resulted in the undesirable stabilisation of the no-scale modulus and, hence of $m_{3/2}$, at the fermionic point. This was the case for the model studied in \cite{Angelantonj:2006ut} albeit with different motivation, which is one of the very few tachyon-free examples in the literature where the one-loop effective potential was investigated in detail around self-dual points.

The dominant asymptotics can be re-expressed in a form that is manifestly invariant under the $\Gamma_0(2)_U$ T-duality group factor as
\begin{align}
	I_{\rm deg}[^0_1] &\simeq \frac{4\zeta(6) c[^0_1](0,0)}{\pi^3 T_2^2} \left[ 2^5 E(3;2U)-E(3;U)\right] \,,
	\label{domasymptot}
\end{align}
in terms of  the real analytic Eisenstein series
\begin{align}
	E(s;\tau) \equiv \frac{1}{2}\sum_{m,n\in\mathbb Z\atop (m,n)\neq(0,0)}\frac{\tau_2^s}{|m+\tau n|^{2s}} \,.
\end{align}
Evaluating  eq. \eqref{domasymptot} at $T=-1+iT_2$ and $U=(1+i)/2$, we can extract the approximate order of magnitude of the potential (in string units), in the large volume limit
\begin{align}
	V_{\rm one-loop}(T_2) \simeq +\frac{0.1727}{T_2^2}\,.
	\label{domasymptot2}
\end{align}
\begin{figure}[t]
	\centering
    \includegraphics{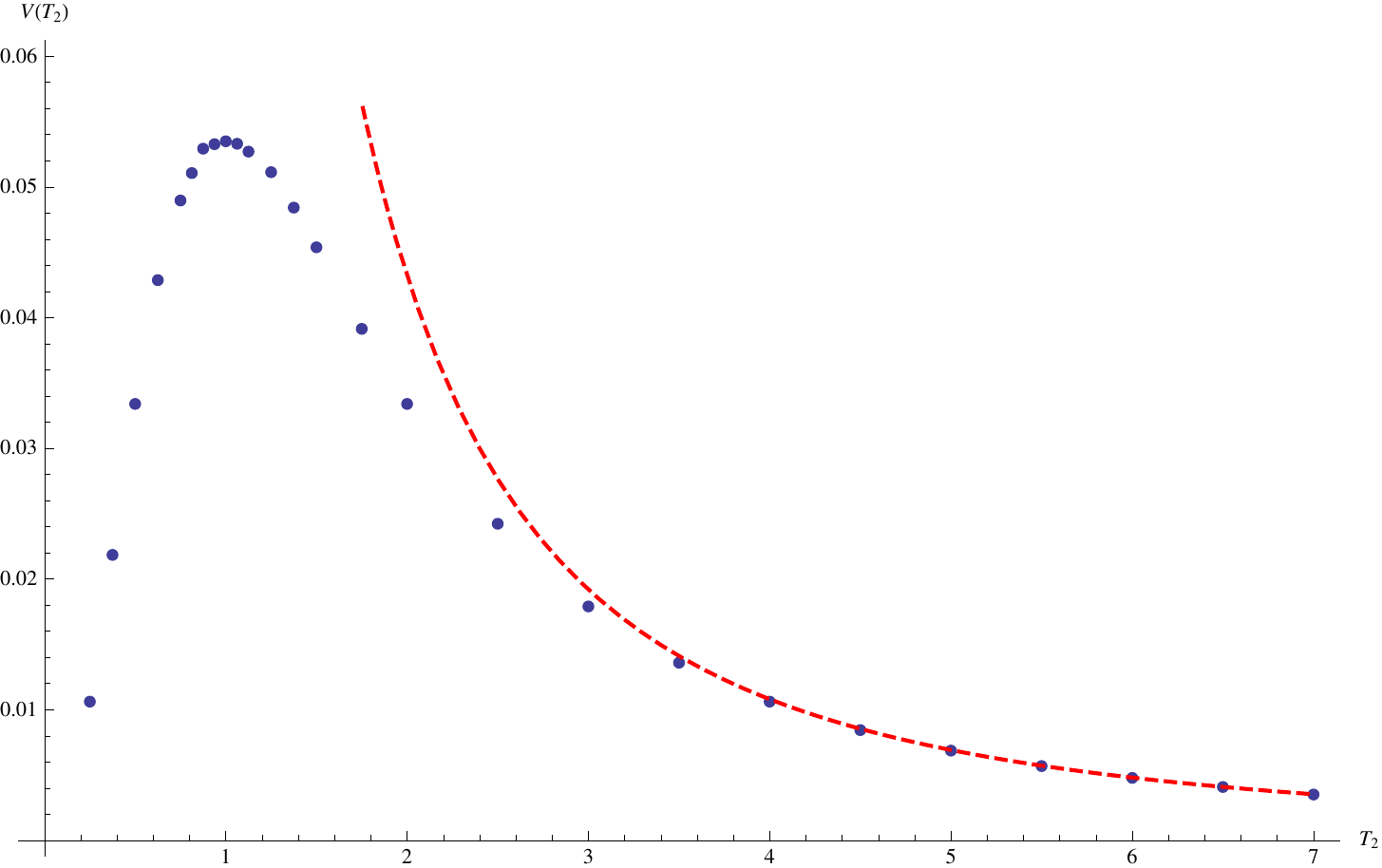}
    \caption{{\it The asymptotic form of the one-loop potential  versus the modulus $T_2$ (dashed line) in Model A matched against the direct numerical evaluation of the integral  \eqref{potentialdef} without unfolding  (in dots). The asymptotic form deviates as one approaches the maximum due to convergence issues arising from the exponential growth of string states.}}\label{themodel}
\end{figure}

Our full result for the one-loop effective potential as a function of $T_2$ is plotted in Figure \ref{themodel}. The numerical evaluation of the potential (dots)  correctly exhibits the T-duality symmetry under $T_2\to 1/T_2$, with a maximum at the self-dual point, and a runaway structure dynamically leading to large values for the volume modulus. As anticipated, the asymptotic form \eqref{domasymptot} of the potential begins to deviate from the numerical one, as we approach the fermionic point.  

It should be stressed that the subleading contribution in the second line of \eqref{AsymptoticExpr}, originating from the massive modes, is much more subtle. The suppression arising from the Bessel factor is considerably smaller, falling only by a factor $\sim e^{- \pi \sqrt{2T_2}}$. This has to balance the highly oscillating, sign-alternating, and exponentially growing (in magnitude) degeneracy of massive level-matched states, entering the coefficients $c[^0_1](N,0)$. As discussed above, the result of this delicate balance yields only an asymptotic expression valid for large $T_2\gg 1$. Its contribution can be estimated by comparing the the deviation of the asymptotic $1/T_2^2$ behaviour, from the direct numerical evaluation of the potential without unfolding the integral.

Although this toy model exhibits some of the attractive features that one would expect from a phenomenologically viable model, it also suffers from certain  serious setbacks. One of them is a question of scales. Assuming that the roll of the no-scale modulus down the one-loop potential and, hence, the scale of supersymmetry breaking $m_{3/2}$, is eventually stabilised by some non-perturbative mechanism in the TeV range, the value of the cosmological constant naively predicted by eq. \eqref{domasymptot2} is huge - overshooting the observed value by some 34 orders of magnitude.

This is actually a generic characteristic of models with a one-loop potential that asymptotically falls power-like, as $\sim 1/T_2^2$. An  inspection of \eqref{AsymptoticExpr} suggests an obvious  way to remedy this situation. The idea is to eliminate the power-like suppression by constructing models with an equal number of massless bosons and fermions at the generic point in the $(T,U)$ moduli space, $c[^0_1](0,0)=0$. In that case, one expects an exponential suppression of the magnitude of the vacuum energy, arising from the subleading term given in the second line of eq. \eqref{AsymptoticExpr}.  Within this context, this possibility was recently considered in \cite{Abel:2015oxa}  and the corresponding models were termed `super no-scale' in \cite{Kounnas:2016gmz}, 
where, however, only the non-chiral case was considered.


\section{Model B : a super no-scale counter-example}\label{counterexamplemodel}

As discussed above, by imposing the condition $n_B=n_F$ on the states which are massless at the generic point in the perturbative $(T,U)$ moduli space, one may indeed achieve an exponentially fast suppression of the cosmological constant. However, although being necessary in order to ensure the small value of the effective potential at large volume, the simple condition $n_B=n_F$ is by no means sufficient on its own. Namely, it does not ensure that the one-loop potential be globally well-behaved.

As we announced already in previous sections, the contribution of extended symmetry points may significantly alter the morphology of the effective potential and spoil its initial run-away behaviour. This can be the case, for example,  if there is an abundance of massless bosons over fermions at the fermionic point. We will illustrate this point by considering a specific counter example: a chiral super no-scale model with $n_B=n_F$ for the massless states at a generic value of $T_2\neq 1$, but with $n_B>n_F$ for the massless spectrum at $T_2=1$, leading to a negative value for the cosmological constant, and stabilising the no-scale modulus at the fermionic point.

The model under consideration, which we henceforth refer to as `Model B', has $N=8$ net matter generations, defined in the fermionic construction by means of the following GGSO matrix
\begin{align}
	c_{\rm(B)}[^{\beta_i}_{\beta_j}] =\left(
\begin{array}{ccccccccc}
 1 & 1 & 1 & -1 & 1 & 1 & 1 & 1 & -1 \\
 1 & 1 & 1 & 1 & -1 & 1 & 1 & 1 & -1 \\
 1 & 1 & -1 & -1 & 1 & -1 & 1 & -1 & 1 \\
 -1 & 1 & -1 & 1 & 1 & 1 & -1 & -1 & 1 \\
 1 & -1 & 1 & 1 & -1 & 1 & -1 & -1 & -1 \\
 1 & -1 & -1 & 1 & 1 & 1 & 1 & -1 & 1 \\
 1 & -1 & 1 & -1 & -1 & 1 & 1 & -1 & -1 \\
 1 & 1 & -1 & -1 & -1 & -1 & -1 & 1 & 1 \\
 -1 & -1 & 1 & 1 & -1 & 1 & -1 & 1 & -1
\end{array}
\right) \,.
\end{align}
As was the case with Model A of section \ref{examplemodel}, also Model B was obtained in our computer-aided scan and, therefore, it satisfies the criteria for the absence of physical tachyons at the fermionic point, as well as the requirement of admitting an interpretation as a spontaneous breaking of supersymmetry by means of a Scherk-Schwarz shift along the first 2-torus.

In addition, it also satisfies the condition $n_B=n_F$ for the massless states at the generic point, while at the fermionic point its $q$-expanded partition function reads
\begin{align}
	\begin{split}
	Z= 8+1760 q+\frac{2}{\bar{q}}+\frac{56 q}{\bar{q}}-\frac{32 q^{1/4}}{\bar{q}^{3/4}}+\frac{224 \sqrt{q}}{\sqrt{\bar{q}}}-\frac{1024 q^{3/4}}{\bar{q}^{1/4}}+1984 q^{1/4} \bar{q}^{1/4}+30720 \sqrt{q} \sqrt{\bar{q}} 
	 +\frac{2048 \bar{q}^{3/4}}{q^{1/4}}+\ldots
	\end{split}
\end{align}
From this expression it is clear that, at the extended symmetry point, the massless spectrum  is dominated by bosons, $n_B-n_F=8$. A naive continuity argument would then suggest that the potential exhibits the form of a `puddle' around the fermionic point, with a negative value for the cosmological constant at the minimum, while rising exponentially fast towards zero as soon as we deform to larger volume. We will see below that this is indeed the case.

The structure of the partition function is precisely identical to the one given in eq. \eqref{orbifoldpartitionf}, the only difference being in the choice of the modular covariant phase, which is now given by
\begin{align}
	\begin{split}
	\Phi = &  ab+ a g_2+b h_2+h_2 g_2 \\
			& +k g_1+\ell h_1 + h_1 g_1\\
			& +\rho G+\sigma H+ HG \\
			& +H_1(b+\ell+\sigma)+G_1(a+k+\rho)+H_1 G_1\\
			& +H_1 g_2+G_1 h_2\\
			&+ H_1 G_2 + G_1 H_2\\
			& +H_2 g_1+G_2 h_1\\
			& +H_2 g_2+G_2 h_2\\
			& +H_3 g_1+G_3 h_1\\
			& + H_3 G+ G_3 H \,.
	\end{split}
\end{align}

This model shares the same generic features as the one discussed in section \ref{examplemodel}. For example, the gravitino mass term, the gauge group at the generic point, the residual T-duality group for each 2-torus and the fact that the vacuum energy integrand receives non-trivial contributions only from the sectors $h_1=g_1=0$, are identical to those discussed in the previous section for model A, and the corresponding arguments shall not be repeated here.

We proceed now to give the formal definition of the model as a $T^6 / (\mathbb{Z}_2)^6$ orbifold, where again  the six $\mathbb Z_2$ factors are associated with their corresponding boundary condition parameters as in eq. \eqref{z6param}. The action on the worldsheet degrees of freedom is now given by
\begin{align}
	\begin{split}
		& \mathbb Z_2^{(1)} \ : \ X^{1,2,5,6}\to - X^{1,2,5,6}\\
		& \mathbb Z_2^{(2)} \ : \ X^{3,4,5,6}\to - X^{3,4,5,6}\\
		& \mathbb Z_2^{(3)} \ : \ (-1)^{F_{\rm s.t.}+F_1+F_2}\,\delta_1 \quad,\ \delta_1: \{ X_1\to X_1+\pi R_1\} \\
		& \mathbb Z_2^{(4)} \ : \  X_3\to X_3+\pi R_3 \\
		& \mathbb Z_2^{(5)} \ : \  X^5\to X^5+\pi R_5 \\
		& \mathbb Z_2^{(6)} \ : \ \bar\phi^{5,6,7,8}\to -\bar\phi^{5,6,7,8}\,,
	\end{split}
\end{align}
together with the following choice of non-trivial discrete torsion 
\begin{align}
	\epsilon(1,4),\ \epsilon(1,5),\ \epsilon(2,3),\ \epsilon(2,4),\ \epsilon(3,4),\ \epsilon(5,6)\,.
\end{align}
\begin{figure}[t]
	\centering
    \includegraphics[width=\textwidth]{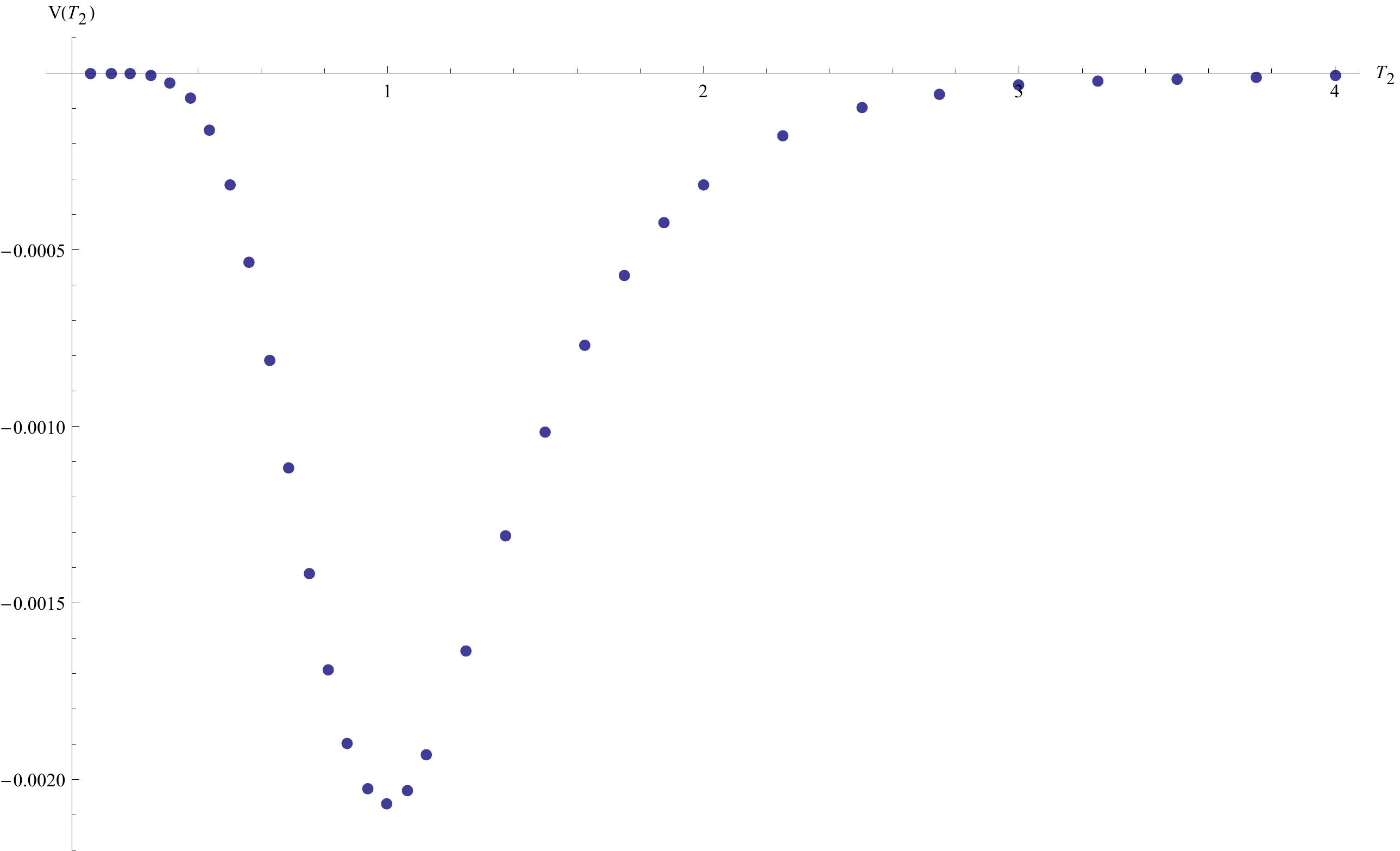}
    \caption{{\it The 
one-loop potential of Model B as a function of $T_2$, obtained by direct numerical evaluation of the integral  \eqref{potentialdef} without unfolding.}}\label{counterexampleFig}
\end{figure}

As in the previous section, we again consider the simple case where all moduli are frozen at their fermionic values as in \eqref{modulideformation}, and deform only with respect to the volume modulus $T_2$ of the first 2-torus. For the purposes of evaluating the one-loop potential, we similarly use the Riemann identity \eqref{Riemannident} to perform the sum over the $(a,b)$ spin structures and restrict our attention to the non-vanishing sectors $h_1=g_1=0$ and $(H_1,G_1)\neq(0,0)$. The integrand can be cast in the simplified form of eq. \eqref{Zsimplified}, with the phase $\hat \Phi$ being given by
\begin{align}
	\begin{split}
	\hat \Phi  = &  HG+(\gamma_2+k+\rho)G_1+(\delta_2+\ell+\sigma)H_2 \\
	 			& +(\gamma_2 g_2 + \delta_2 h_2)\\
				& + (\gamma_3+\rho)G+(\delta_3+\sigma)H \\
				& + g_2(1+H_1+k+h_2) \,.
	\end{split}
\end{align}
From this form, it is possible to unfold the modular integral and obtain the asymptotic expansion of the one-loop potential. The result is again of the generic form \eqref{AsymptoticExpr},  but the first line proportional to $c[^0_1](0,0)=n_B-n_F=0$ is now absent by construction. As desired, the dominant suppression of the magnitude of the cosmological constant originates from the Bessel series in the second line of \eqref{AsymptoticExpr} and indeed becomes exponentially small for large $T_2\gg 1$.

However, this does not imply that the model dynamically leads to a large volume scenario and, therefore, to supersymmetry breaking at low scales. In fact, the structure of the potential is significantly affected by the abundance of extra massless bosons arising at $T_2=1$. The situation is best illustrated by the numerical evaluation of the one-loop potential given in Figure \ref{counterexampleFig}. As anticipated, it  has  the shape of a puddle with a minimum at the fermionic point $T_2=1$, giving rise to a large negative value for the cosmological constant. 

What this simple toy model illustrates is that, even though the super no-scale condition $n_B=n_F$ is satisfied by massless states at the generic point in the $(T,U)$ moduli space, it is certainly not sufficient to determine the attractive or repulsive form of the potential around the self-dual points. Moreover, due to the attractor nature of the potential leading to the stabilisation of the no-scale modulus $T_2$ at the self-dual point, the question of tree-level stability becomes relevant once again, because of the possibility of exciting tachyonic modes as soon the Scherk Schwarz radius approaches the string scale. A much more surprising situation will be discussed in the following sections.


\section{Conditions for small and positive cosmological constant}\label{conditionsmodels}

Thus far, in our discussion based on models A and B, we have seen two simple conditions on the massless string spectra that appeared to govern the form of the one loop effective potential. On the one hand, the generic asymptotic behaviour \eqref{AsymptoticExpr} suggests that the super no-scale requirement $n_B=n_F$ for a degeneracy in the number of massless bosons and fermions at a generic point in the $(T,U)$ moduli space, is necessary in order to guarantee an exponentially small value for the cosmological constant at large volume. 

On the other hand, we illustrated in the previous section that this alone is not sufficient to fix the global shape of the potential. In both models A and B, the  additional states becoming massless at self-dual points were seen to crucially affect the nature of the local extremum about these points. Indeed, Model A exhibited $n_B<n_F$ both at the fermionic as well as the generic point and its potential was characterised by positive values and a  maximum occurring at the fermionic point $T_2=1$.   Model B instead satisfied $n_B=n_F$ at the generic point, but with $n_B>n_F$ at the fermionic point, and its potential had the form of a puddle, with a minimum at negative values, stabilising the SUSY breaking scale at the string scale.

A natural possibility would be to try to combine the virtues of models A and B, and search for constructions with super no-scale structure, $n_B=n_F$ at the generic point, while requiring an abundance of fermions at the fermionic point. Indeed, one could expect the condition $n_F>n_B$ at the fermionic point to induce a potential in the shape of a bump, with a maximum located at $T_2=1$ at positive values, while simultaneously imposing $n_B=n_F$ at the generic point would additionally lead to a fast decay to exponentially small (but still positive) values for the cosmological constant.

By  exploiting the origin of this class of models as deformations away from the point $T_2=1$, where a formulation in terms of free fermions exists, it was possible to derive explicit algebraic conditions in terms of the GGSO coefficients $c[^{\beta_i}_{\beta_j}]$ that guarantee the super no-scale properties of the theory. Although the precise  mathematical expression for these conditions is technically  involved and will not be given explicitly in the present work, it was nevertheless fully incorporated in our computer-aided scan. Unfortunately, our random scan of a sample of $10^6$ models did not produce a single model satisfying both conditions. Of course, this is by no means a general no-go statement, since it is only based on a random scan and on the specific choice of basis vectors \eqref{basis}.

Nevertheless, at first sight, this discouraging fact might seem to imply that the construction of heterotic chiral super no-scale models,  with a positive potential in the form of a bump leading to a dynamical roll of the Scherk-Schwarz volume $T_2$ away from the extended symmetry points, might not be possible. Fortunately, this is not the case. The morphology of the one-loop effective potential in the class of models under consideration is actually much richer than one might naively expect from a cursory inspection of the massless sector. Aside from massless modes, string amplitudes receive contributions not only from the infinite massive towers of oscillator, Kaluza-Klein and winding states, but also from non-level matched states. Although the contribution of the latter is washed out after one performs the $\tau_1$ integral in the field theory limit $\tau_2\to\infty$, where the fundamental domain $\mathcal F$ becomes approximately rectangular, such unphysical string states can have non-trivial contributions arising from the curved region of $\mathcal F$ at $\tau_2 \leq 1$. 

As we shall see explicitly in this section, contrary to our field-theoretic intuition, it is precisely the effect of these non level-matched states that enables the construction of heterotic vacua with large positive cosmological constant that may naturally decay into a vacuum with exponentially small cosmological constant and gravitino mass scale which can be in the TeV range.

Let us begin by considering more carefully the contributions to the one-loop potential arising from the states of a generic fermionic model, in the class we are studying. We will expand its partition function in powers of $q_{\rm r}=e^{-2\pi \tau_2}$ and $q_{\rm i}=e^{2\pi i \tau_1}$ as follows
\begin{align}
	Z= \sum_{n\in \mathbb Z/2\atop n\geq -1/2}\sum_{m\in\mathbb Z} Z_{n,m}\,q_{\rm r}^n \,q_{\rm i}^m \,.
	\label{Zexpand}
\end{align}
The coefficients  $Z_{n,m}$ are not arbitrary but are constrained by modular invariance to be integer numbers, and highly depend on the particular model under consideration. They describe the degeneracy between bosonic and fermionic excitations at mass level $n$ and conformal weights $(\frac{n+m}{2},\frac{n-m}{2})$, as required by the particle interpretation of $Z$. Properly incorporating the left- and right- moving ground states of the heterotic string worldsheet at the fermionic point implies $n\geq -1/2$ and increasing in steps of half a unit of conformal weight, whereas $m$ is constrained by modular invariance under $\tau\to\tau+1$ to be an integer in the range $-[n]-1\leq m\leq [n]+2$, where $[n]$ denotes the integer part of $n$. 

It is useful for the subsequent discussion to view \eqref{Zexpand} as an expansion in powers of $q_{\rm r}$, with coefficients which are in turn expanded in powers of $q_{\rm i}$
\begin{align}
	Z = \sum_{n\in\mathbb Z/2\atop n\geq -1/2} \left[ \sum_{ m=-[n]-1}^{[n]+2} Z_{n,m}\, q_{\rm i}^m \right]\,q_{\rm r}^n \,.
	\label{Zexpand2}
\end{align}
 It is straightforward to see that for a given mass level $n$, there is a finite number of $q_{\rm i}$-terms, with the $m=0$ mode corresponding to the physical (level-matched) string excitations. This arrangement of the expansion has the advantage that it groups together terms with comparable contribution to the one-loop potential. Indeed, terms at the same order of  $q_{\rm r}$ yield similar order of magnitude contributions to the integral, while terms of higher order $n$  are exponentially suppressed.

It is now convenient to partition the integration domain into two regions $\mathcal F = S_1 \cup S_2$, such that $S_1$ is rectangular and contains the infrared point $\tau=i\infty$
\begin{align}
	S_1 = \{ \tau\in\mathcal F \,|\, \tau_2>1 \} \,,
\end{align}
while $S_2$ is the curved region
\begin{align}
	S_2 = \{ \tau\in \mathcal F \, | \, \tau_2 <1 \} \,.
\end{align}
Although only level matched states $m=0$ may contribute to the integral over $S_1$ thanks to the integration over $x\equiv {\rm Re}(\tau)$, this is not the case with the integral over $S_2$. Plugging the expansion \eqref{Zexpand2} directly into the integral \eqref{potentialdef} for the one-loop potential, and performing the splitting we have 
\begin{align}
	2(2\pi)^4\,V_{\rm 1-loop}= \sum_{n}Z_{n,0} \,I_{n,0}^1 +\sum_{n,m}Z_{n,m}\, I_{n,m}^2 \,,
	\label{potexpansion}
\end{align}
where $I_{n,m}^1$ and $I_{n,m}^2$ are the corresponding integrals in regions $S_1$, $S_2$, respectively,
\begin{align}
	\begin{split}
	I_{n,m}^1 &\equiv -\delta_{m,0}\,\int_{1}^{\infty} dy\,\frac{e^{-2\pi n y}}{y^3} = -\delta_{m,0}\, (2\pi n)^2\,\Gamma(-2, 2\pi n)\,, \\
	I_{n,m}^2 &\equiv -\int_{-1/2}^{1/2} dx\,e^{2\pi i |m| x}\,\int_{\sqrt{1-x^2}}^{1} dy\,\frac{e^{-2\pi n y}}{y^3} \,.
	\end{split}
\end{align}
The explicit evaluation of $I_{n,m}^1$ presents no difficulty, since the rectangularity of $S_1$ implies that the two-dimensional integral over $x\equiv {\rm Re}(\tau)$ and $y\equiv {\rm Im}(\tau)$ splits into a product of two, independent, one-dimensional ones. Indeed,  the $x$-integration simply imposes level matching and projects onto the $m=0$ sector, while the integral over $y$ is essentially the definition of the incomplete Gamma function. The case $m=n=0$ is also incorporated in the above expression for $I^1$, with the understanding that it is obtained as the formal limit $n\to 0$,  yielding $I_{0,0}^{1} = -1/2$. 

Of course, the positivity of the integrand reflects itself in the fact that  $I_{n,m}^1$ is always negative, which is in perfect agreement with our expectation from field theory. Indeed, as long as one considers only contributions to the potential arising from $I^1$, which involves only the level-matched sector, an abundance of bosons $Z_{n,0}>0$ at some mass level $n$ leads to a negative contribution to the one-loop potential, $Z_{n,0}I_{n,0}^1$. An abundance of fermions $Z_{n,0}<0$ instead produces a positive contribution. 

This simple field-theoretic intuition was our guiding principle in constructing models A, B in the previous sections. There, we focused on the difference $Z_{0,0}$ between the number of massless bosonic and fermionic modes as the primary factor determining the sign of the potential at the fermionic point. Although, this is the case for a variety of models, this has to be weighted against the possibility that non level-matched contributions from $I^2_{n,m}$ might actually reverse the situation and a careful analysis is called for. In the case of $I_{n,m}^1$, it is straightforward to see that it falls exponentially fast at high mass levels $2\pi n\gg 1$ according to the asymptotic formula
\begin{align}
	I_{n,0}^1 \sim -\frac{e^{-2\pi n}}{2\pi n} \,.
\end{align}
Since the degeneracy of states $Z_{n,m}$  can at most grow at most as $\sim e^{c\sqrt{n}}$, where $c$ is some positive constant, it is clear that after the first few orders, the contributions become negligible. 

Let us now consider the more interesting case of $I_{n,m}^2$. As anticipated, this is much more subtle due to the non-rectangular shape of $S_2$ and will receive non-vanishing contributions also from non-level matched modes, $m\neq 0$. Let us consider first the massless case $n=0$. For the level-matched massless states the integral $I_{0,0}^2$ may be evaluated analytically and gives a negative contribution $I_{0,0}^2 = -(\log 3-1)/2\simeq -0.049$. For massless but non-level matched states, $I_{0,m}^2$  may be  expressed in terms of the incomplete Gamma function as
\begin{align}
	I_{0,m}^2 =\frac{1}{4} \left[ \Gamma(0,3 \pi i |m|)-\Gamma(0, \pi i |m|) + \Gamma(0,-3\pi i |m|)-\Gamma(0,-\pi i |m|)  \right]\,,
\end{align}
and for large $|m|\gg 1$, its highly oscillatory behaviour is governed by the asymptotic expression
\begin{align}
	I_{0,m}^2 \sim (-1)^{m+1} \left(\frac{2}{3\pi m}\right)^2 \,.
\end{align}
However, due to the structure of the expansion, only the integrals $I_{0,1}^2\simeq 0.031$ and $I_{0,2}^2\simeq -0.01$ may actually appear as contributions to the potential, illustrating the general principle that both signs typically appear at comparable orders of magnitude as long as $n$ is kept fixed and $m$ is varied. 

Next, consider the more interesting case of massive states $n>0$. For the level-matched ones, it is possible to find the following asymptotic expansion
\begin{equation}
	\begin{split}
	I_{n,0}^2 &= 2i\sqrt{2}e^{-2\pi n}\sum_{k,\ell\geq 0}\frac{(-1)^{\ell}\,(2k)!(\ell+1)(\ell+2)}{2^{3k+1}k!(2\pi n)^{k+\ell+3/2}} \,{\rm Im}\left[ \Gamma(k+\ell+\tfrac{3}{2}, n\pi(-2+\sqrt{3}))-\Gamma(k+\ell+\tfrac{3}{2})\right]\\
		& -(2\pi n)^2 \left[\Gamma(-2,\pi n\sqrt{3})-\Gamma(-2,2\pi n)\right] \,,
	\end{split} 
\end{equation}
which provides a very good approximation to the integral even if only the first few terms in the sum over $k,\ell$ are kept. The dominant asymptotic behaviour can be, hence, extracted from the term $k=\ell=0$ as follows
\begin{align}
	I_{n,0}^2  \sim -\frac{4}{3}\,\frac{e^{-\pi n \sqrt{3}}}{\pi n \sqrt{3}}\left[1- \frac{3\sqrt{3}}{4\sqrt{2}}(\sqrt{3}-1)\left(1-\frac{2+\sqrt{3}}{2\pi n} \right)\right] + \frac{e^{-2\pi n}}{2\pi n } \,.
\end{align}
From this asymptotic form we see that, again, level matched states $I_{n,0}^2$ from region $S_2$ still produce a negative contribution, as expected, and that higher mass levels are exponentially suppressed by at least a factor of $e^{-\pi n \sqrt{3}}$.

The last remaining case to consider is $I_{n,m}^2$ containing the contribution of non level-matched states, $m\neq 0$, at various non-vanishing mass levels, $n\neq 0$. Deriving an exact formula for this integral can be quite tedious due to the presence of effectively two scales in the problem, set by the integers $n$ and $m$, controlling the exponential suppression and oscillation, respectively. Although it is possible to obtain certain exact series representations with fast convergence around various limits for $(n,m)$, their precise mathematical form is quite complex and we will not display them here explicitly. Instead, it will be simpler for our arguments to consider an approximate asymptotic formula that captures the essential behaviour of the integral in the parameter range of interest
\begin{align}
	I_{n,m}^2 \sim \frac{(-1)^{m+1}}{2(\pi m)^2} \left(1-\frac{1}{2m}\right) \frac{ e^{-2\pi n \sqrt{1-\frac{1}{4}\left(1-\frac{1}{2m}\right)^2}}}{ \left[ 1-\frac{1}{4}\left(1-\frac{1}{2m}\right)^2\right]^2} \,.
\end{align}
Already this approximate formula, which is valid even for $n<0$ as long as $m\neq 0$, illustrates quite clearly that that non-negligible positive contributions do arise from non-level matched states with odd $m$.

\begin{table*}\centering
\ra{1.3}
\begin{tabular}{@{}r|rrrrr@{}}\toprule
$n $ & $m$ &  $0$ & $\pm 1$ & $\pm 2$ & $\pm 3$ \\ \midrule
$-1$ &&  N/A & 0 & N/A & N/A \\
$-\frac{1}{2}$ &&  N/A & 0 & N/A & N/A \\
0 && $-0.500$ & 0 & 0 & N/A \\
$\frac{1}{2}$ && $-0.00755$ & 0 & 0 & N/A \\
1 && $-0.000208$ & 0 & 0 & 0 \\
$\frac{3}{2}$ && $-6.61\times 10^{-6}$ & 0 & 0 & 0 \\ \bottomrule
\end{tabular}
\caption{{\it Numerical values of the integral $I_{n,\pm m}^1$ for the first few energy levels.} }\label{tableI1}
\end{table*}
\begin{table*}\centering
\ra{1.3}
\begin{tabular}{@{}r|rrrrr@{}}\toprule
$n $ & $m$  & $0$ & $\pm 1$ & $\pm 2$ & $\pm 3$ \\ \midrule
$-1$&&  N/A & 12.2 & N/A & N/A \\
$-\frac{1}{2}$ &&  N/A & 0.617 & N/A & N/A \\
0 &&  $-0.0493$ & 0.0315 & $-0.00989$ & N/A \\
$\frac{1}{2}$ &&  $-0.00245$ & 0.00163 & $-0.000587$ & N/A \\
1 &&  $-0.000123$ & 0.0000846 & $-0.0000346$ & 0.0000180 \\
$\frac{3}{2}$ &&  $-6.24\times 10^{-6}$ & $4.45\times 10^{-6}$ & $-2.02\times 10^{-6}$ & $1.11\times 10^{-6}$ \\ \bottomrule
\end{tabular}
\caption{{\it Numerical values of the integral $I_{n,\pm m}^2$ for the first few energy levels.} }\label{tableI2}
\end{table*}

The integrals $I_{n,m}^1$ and $I_{n,m}^2$ are model independent quantities that play a central role in organising the various contributions to the one-loop potential. To illustrate these contributions and see how the various states of models A and B contributed to its shape at the fermionic point, we have tabulated their explicit numerical values for the first few mass levels in Tables \ref{tableI1} and \ref{tableI2}, with precision of 3 significant figures. Entries marked as N/A in these tables are irrelevant to our analysis, since the corresponding terms in the expansion \eqref{potexpansion} do not occur  in the class of models under consideration.

An inspection of Tables \ref{tableI1} and \ref{tableI2}, at each mass level, reveals that not only do the non level-matched states contribute to the one-loop potential with comparable magnitude to the level-matched ones, but that this contribution might in fact be sizeable and positive, \emph{e.g.}  $I^2_{-1,\pm 1} \simeq 12.2 $. With these values at our disposal, we are now ready to perform the anatomy of the contributions of the various physical and unphysical states to the one-loop potential in models A and B. We begin by expanding the corresponding partition functions according to \eqref{Zexpand2},
\begin{align}
	\begin{split}
		Z_{(A)} &= \frac{2q_{\rm i}}{q_{\rm r}}-\frac{16 q_{\rm i}}{\sqrt{q_{\rm r}}}+(-312+32q_{\rm i}+56 q_{\rm i}^2)+\left(4064+\frac{6144}{q_{\rm i}}+512q_{\rm i}-416q_{\rm i}^2\right)\sqrt{q_{\rm r}} \\
			    &+\left(12288+\frac{16384}{q_{\rm i}^2}+\frac{103680}{q_{\rm i}}-12320q_{\rm i}-256q_{\rm i}^2+792q_{\rm i}^3\right)q_{\rm r}  \\
			    & +\left( -537600+\frac{790528}{q_{\rm i}^2}+\frac{892976}{q_{\rm i}}+101568q_{\rm i} +11264q_{\rm i}^2-5520q_{\rm i}^3\right) q_{\rm r}^{3/2} + \ldots
	\end{split}
\end{align}
\begin{align}
	\begin{split}
		Z_{(B)} &=\frac{2 q_{\rm i}}{q_{\rm r}}-\frac{32 q_{\rm i}}{\sqrt{q_{\rm r}}}+\left(8+224 q_{\rm i}+56 q_{\rm i}^2\right)+\left(1984+\frac{2048}{q_{\rm i}}-1024 q_{\rm i}-832 q_{\rm i}^2\right) \sqrt{q_{\rm r}}\\ 
		&+\left(30720+\frac{10240}{q_{\rm i}^2}+\frac{92160}{q_{\rm i}}+1760 q_{\rm i}+5376 q_{\rm i}^2+792 q_{\rm i}^3\right) q_{\rm r}\\
		&+\left(-395264+\frac{569344}{q_{\rm i}^2}+\frac{1003616}{q_{\rm i}}-54912 q_{\rm i}-22528 q_{\rm i}^2-11040 q_{\rm i}^3\right) q_{\rm r}^{3/2} +\ldots
	\end{split}
\end{align}

\begin{table*}\centering
\ra{1.3}
\begin{tabular}{@{}r|rrrr@{}}\toprule
$n $ & Model  & A && B  \\ \midrule
$-1$&&  24.4 && 24.4 \\
$-\frac{1}{2}$ &&  $-9.87$ && $-19.7$ \\
0 &&  $172.$ && 2.11 \\
$\frac{1}{2}$ &&  $-29.6$ && $-17.7$  \\
1 &&  $3.13$ && $-2.73$  \\
$\frac{3}{2}$ &&  $9.71$ && $8.18$  \\ \midrule
 Total && +170. && $-5.47$ \\ \bottomrule
\end{tabular}
\caption{{\it Contributions to the rescaled one-loop potential $2(2\pi)^4 V_{\rm 1-loop}$ arranged according to energy level for models A and B. At each level $n$, the cumulative contribution of level-matched as well as non level-matched states is displayed.} }\label{tableNumAB}
\end{table*}
Plugging the above expansions together with the numerical values of the integrals $I_{n,m}^1$ and $I_{n,m}^2$ into \eqref{potexpansion}, we can see how the various states contribute to the determination of the sign of the effective potential at the fermionic point. The numerical contributions to the (rescaled) potential $2(2\pi)^4 V_{\rm 1-loop}$ for models A and B are presented in Table \ref{tableNumAB} for the first few energy levels. We would like to stress that although for simplicity we only explicitly display the contributions up to mass level $n=3/2$, the results remain essentially unaltered as the order increases.

An inspection of Table \ref{tableNumAB} shows already that the contribution of the necessarily non level-matched, negative mass levels $n=-1$ and $n=-1/2$ is significant for both models. Subsequently, at the massless level $n=0$, model A is dominated by the contribution of the abundance of 312 massless fermionic states, which gives rise to an enormous contribution $\sim 172$ to the integral. That this huge number can actually arise is possible precisely because of the very fact that model A is not constrained by the super no-scale conditions at the generic point in $(T,U)$ moduli space, and the situation remains largely unchanged as massive levels $n>0$ are taken into account. 

On the contrary, model B exhibits only a modest contribution $\sim +2.11$ at $n=0$, out of which only $\sim -4.39$ is due to the level-matched massless states. It is very interesting that, up until the massless level $n\leq 0$, unphysical state contributions completely dominate over those of physical ones and, in fact, summing up all contributions up to $n=0$ would have given rise to a positive net value $\sim +6.81$ for the one-loop potential. The situation changes drastically as soon as the first few massive levels $n>0$ are considered. Indeed, for $n>0$, the most significant contributions are negative and arise from massive level-matched states, which add up to eventually generate a negative total value $\sim -5.47$ for the (rescaled) one-loop potential.

What this anatomy of the energy budget of models A and B teaches us is that the shape of the stringy one-loop potential around self-dual points can be quite intricate. Its form may be significantly affected by both level-matched as well as non-level matched states, including so-called unphysical tachyons ($m\neq 0$ and $n<0$).  As a result of this investigation and in what concerns model building, we may already conclude that any argument or condition imposed on the string spectrum for purposes of controlling the form of the one-loop potential around self-dual points, should  not  be restricted to  the massless physical sector alone. Instead, it should necessarily take into account both the non level-matched as well as the first few massive states.

Therefore, the requirement of positivity of the potential at the fermionic point translates itself into constraints for the model dependent coefficients $Z_{n,m}$, such that
\begin{align}
	 \sum_{n}Z_{n,0} \,I_{n,0}^1 +\sum_{n,m}Z_{n,m}\, I_{n,m}^2  >0  \,,
	 \label{conditionRefined}
\end{align}
with the model-independent coefficients $I_{n,m}^1$ and $I_{n,m}^2$ given in Tables \ref{tableI1} and \ref{tableI2}, respectively.
It is in fact possible to satisfy this condition, even if  the low lying level-matched states satisfy $I_{n,0}<0$, depending on the choice of model. 
In the next section, we present a particular example model, which indeed starts with a positive value of the potential at the fermionic point, and then naturally rolls down to attain exponentially small values as soon as the Scherk-Schwarz volume grows sufficiently far away from $T_2=1$.


\section{Model C : a model with small positive cosmological constant}\label{exmetastable}

As discussed in the previous section, the requirement for a positive runaway potential that dynamically leads to large Scherk-Schwarz volume, while maintaining exponentially small values for the cosmological constant in the $T_2\gg 1$ regime, requires the super no-scale property $n_B=n_F$ at the generic point of the $(T,U)$ moduli space, together with the refined condition \eqref{conditionRefined} at the fermionic point. The latter ensures the positivity requirement of the potential and replaces the naive condition for an abundance of fermions in the massless physical spectrum at $T_2=1$. Although imposing \eqref{conditionRefined} analytically seems difficult, it is nevertheless straightforward to implement as an additional constraint in our computer-aided  scan using the fermionic construction, since the latter provides an explicit expression of the coefficients $Z_{n,m}$ directly in terms of the GGSO coefficients $c[^{\beta_i}_{\beta_j}]$.

In this section, we shall present a specific solution to these constraints that we shall refer to as `Model C'. It is a construction with net chirality $N=8$, although very similar constructions exist also for $N=4$ and $N=12$ with an identical form for the one-loop potential.
In its free fermionic realisation, Model C  is defined by the following choice of GGSO matrix
\begin{align}
	c_{\rm(C)}[^{\beta_i}_{\beta_j}] =\left(
\begin{array}{ccccccccc}
 1 & 1 & 1 & 1 & -1 & 1 & 1 & 1 & -1 \\
 1 & 1 & 1 & -1 & -1 & 1 & 1 & 1 & -1 \\
 1 & 1 & -1 & 1 & 1 & -1 & 1 & -1 & 1 \\
 1 & -1 & 1 & -1 & -1 & 1 & -1 & -1 & -1 \\
 -1 & -1 & 1 & -1 & 1 & 1 & -1 & -1 & 1 \\
 1 & -1 & -1 & 1 & 1 & 1 & 1 & 1 & -1 \\
 1 & -1 & 1 & -1 & -1 & 1 & 1 & 1 & -1 \\
 1 & 1 & -1 & -1 & -1 & 1 & 1 & 1 & -1 \\
 -1 & -1 & 1 & -1 & 1 & -1 & -1 & -1 & -1
\end{array}
\right)\,.
\end{align}
As was the case with models A and B, also this model satisfies the criteria of super no-scale structure, of the absence of physical tachyons at the fermionic point, as well as the requirement of admitting a well-defined interpretation as a $T^6/(\mathbb Z_2)^6$ orbifold with  $\mathcal N=1\to 0$ spontaneous breaking of supersymmetry of the Scherk-Schwarz type.

Its partition function at the fermionic point reads 
\begin{align}
	\begin{split}
Z =& +40-10144 q+\frac{2}{\bar{q}}+\frac{56 q}{\bar{q}}+\frac{792 q^2}{\bar{q}}-\frac{16 q^{1/4}}{\bar{q}^{3/4}}-\frac{672 q^{5/4}}{\bar{q}^{3/4}}-\frac{10128 q^{9/4}}{\bar{q}^{3/4}}+\frac{64 \sqrt{q}}{\sqrt{\bar{q}}}+\frac{3072 q^{3/2}}{\sqrt{\bar{q}}} \\
 &+\frac{768 q^{3/4}}{\bar{q}^{1/4}}+\frac{12800 q^{7/4}}{\bar{q}^{1/4}}+224 q^{1/4} \bar{q}^{1/4}-39744 q^{5/4} \bar{q}^{1/4}+14336 \sqrt{q} \sqrt{\bar{q}}+\frac{6912 \bar{q}^{3/4}}{q^{1/4}}\\ 
 & -203776 q^{3/4} \bar{q}^{3/4}+118656 \bar{q}+498224 q^{1/4} \bar{q}^{5/4}+\frac{9216 \bar{q}^{3/2}}{\sqrt{q}}+\frac{934400 \bar{q}^{7/4}}{q^{1/4}}+\ldots \,,
	\end{split}\label{Zmeta1}
\end{align}
and the massless spectrum of bosons and fermions, together with their charges under the ${\rm SO}(10)\times{\rm SO}(8)^2\times{U}(1)^3$ gauge group factors are assembled in Tables \ref{TableSpectrBos} and \ref{TableSpectrFerm}.

\begin{table*}\centering
\ra{1.3}
\begin{tabular}{@{}r|r@{}}\toprule
Sector&${\rm SO}(10)\times{\rm SO(8)}^2\times{\rm U(1)}^3$ representation(s)\\
\midrule
\textbf{1}&$\left({\mathbf{10}},{\mathbf1},{\mathbf1},\pm1,0,0\right)$,
$\left({\mathbf{10}},{\mathbf1},{\mathbf1},0,\pm1,0\right)$,
$\left({\mathbf{10}},{\mathbf1},{\mathbf1},0,0,\pm1\right)$\\
&$\left({\mathbf1},{\mathbf1},{\mathbf1},\pm1,\pm1,0\right)$,
$\left({\mathbf1},{\mathbf1},{\mathbf1},\pm1,0,\pm1\right)$,
$\left({\mathbf1},{\mathbf1},{\mathbf1},0,\pm1,\pm1\right)$\\
&$\left({\mathbf1},{\mathbf1},{\mathbf1},\pm1,\mp1,0\right)$,
$\left({\mathbf1},{\mathbf1},{\mathbf1},\pm1,0,\mp1\right)$,
$\left({\mathbf1},{\mathbf1},{\mathbf1},0,\pm1,\mp1\right)$\\
&$12\times\left({\mathbf1},{\mathbf1},{\mathbf1},0,0,0\right)$\\
\midrule
$T_1+T_3$&
$32\times\left({\mathbf1},{\mathbf1},{\mathbf1},0,0,0\right)$\\
$b_1+x+z_1+T_2+T_3$&
$4\times\left({\mathbf1},{\mathbf8},{\mathbf1},0,+\frac{1}{2},-\frac{1}{2}\right)$
\\
$b_1+x+z_1+T_2$&
$4\times\left({\mathbf1},{\mathbf8},{\mathbf1},0,+\frac{1}{2},-\frac{1}{2}\right)$
\\
$b_1+x$&
$8\times\left({\mathbf1},{\mathbf1},{\mathbf1},+\frac{1}{2},0,+\frac{1}{2}\right)+8\times\left({\mathbf1},{\mathbf1},{\mathbf1},-\frac{1}{2},0,-\frac{1}{2}\right)$
\\
$b_2+x+T_1+T_3$&
$8\times\left({\mathbf1},{\mathbf1},{\mathbf1},\frac{1}{2},0,-\frac{1}{2}\right)+
8\times\left({\mathbf1},{\mathbf1},{\mathbf1},-\frac{1}{2},0,\frac{1}{2}\right)$
\\
$b_2+x+T_1$&
$4\times\left({\mathbf1},{\mathbf8},{\mathbf1},+\frac{1}{2},0,-\frac{1}{2}\right)$
\\
$b_2+x+z_2+T_3$&
$4\times\left({\mathbf1},{\mathbf1},{\mathbf8},+\frac{1}{2},0,+\frac{1}{2}\right)$
\\
$b_2+x$&
$4\times\left({\mathbf{10}},{\mathbf1},{\mathbf1},+\frac{1}{2},0,+\frac{1}{2}\right)+
8\times\left({\mathbf1},{\mathbf1},{\mathbf1},+\frac{1}{2},0,-\frac{1}{2}\right)$\\
&$8\times\left({\mathbf1},{\mathbf1},{\mathbf1},-\frac{1}{2},0,+\frac{1}{2}\right)+4\times\left({\mathbf1},{\mathbf1},{\mathbf1},\frac{1}{2},\pm1,+\frac{1}{2}\right)$
\\
$b_3+x$&
$8\times\left({\mathbf1},{\mathbf1},{\mathbf1},\frac{1}{2},\frac{1}{2},0\right)+
8\times\left({\mathbf1},{\mathbf1},{\mathbf1},-\frac{1}{2},-\frac{1}{2},0\right)$
\\
$b_3+x+T_1+T_2$&
$4\times\left({\mathbf1},{\mathbf1},{\mathbf8},+\frac{1}{2},-\frac{1}{2},0\right)$\\
$b_3+x+T_2+z_2$&
$4\times\left({\mathbf1},{\mathbf1},{\mathbf8},+\frac{1}{2},+\frac{1}{2},0\right)$
\\
$b_3+x+T_1+T_2$&
$4\times\left({\mathbf{16}},{\mathbf1},{\mathbf1},0,0,-\frac{1}{2}\right)$
\\
\bottomrule
\end{tabular}
\caption{\it Spectrum of massless bosonic matter  and quantum numbers under the various gauge group factors. Sectors in the first column are labeled according to the conventions of the fermionic construction for the basis vectors, as given in \eqref{basis}. }\label{TableSpectrBos}
\end{table*}


\begin{table*}\centering
\ra{1.3}
\begin{tabular}{@{}r|r@{}}\toprule
Sector&${\rm SO}(10)\times{\rm SO(8)}^2\times{\rm U(1)}^3$ representation(s)\\
\midrule
$S$&
$4\times\left({\mathbf1},{\mathbf8},{\mathbf1},0,0,0\right)$\\
$S+b_1+T_2$&
$4\times\left(\overline{\mathbf{16}},{\mathbf1},{\mathbf1},-\frac{1}{2},0,0\right)$\\
$S+b_1+x$&
$4\times\left(\mathbf{1},{\mathbf8},{\mathbf1},0,-\frac{1}{2},-\frac{1}{2}\right)$\\
\midrule
$S+b_1+x+T_3$&
$4\times\left(\mathbf{10},{\mathbf1},{\mathbf1},0,-\frac{1}{2},+\frac{1}{2}\right)
+8\times\left(\mathbf{1},{\mathbf1},{\mathbf1},0,\pm1,\frac{1}{2},\pm\frac{1}{2}\right)$
\\
&$4\times\left(\mathbf{1},{\mathbf1},{\mathbf1},\pm,\frac{1}{2},-\frac{1}{2}\right)$
\\
\midrule
$S+b_1+x+z_2+T_2+T_3$&
$4\times\left(\mathbf{1},{\mathbf1},{\mathbf8},0,+\frac{1}{2},+\frac{1}{2}\right)$
\\
$S+b_1+x+T_2$&
$4\times\left(\mathbf{1},{\mathbf1},{\mathbf8},0,-\frac{1}{2},+\frac{1}{2}\right)$\\
$S+b_1+x+z_2$&
$4\times\left(\mathbf{8},{\mathbf1},{\mathbf1},0,+\frac{1}{2},-\frac{1}{2}\right)$\\
$S+b_2+x$&
$4\times\left(\mathbf{1},{\mathbf8},{\mathbf1},0,-\frac{1}{2},+\frac{1}{2}\right)$\\
$S+b_2+x+T_3$&
$8\times\left(\mathbf{1},{\mathbf1},{\mathbf1},+\frac{1}{2},0,-\frac{1}{2}\right)
+8\times\left(\mathbf{1},{\mathbf1},{\mathbf1},-\frac{1}{2},0,+\frac{1}{2}\right)$\\
\midrule
$S+b_2+x+T_1$&
$4\times\left(\mathbf{10},{\mathbf1},{\mathbf1},+\frac{1}{2},0,+\frac{1}{2}\right)
+8\times\left(\mathbf{1},{\mathbf1},{\mathbf1},0,\pm\frac{1}{2},\mp\frac{1}{2}\right)$
\\
&$4\times\left(\mathbf{1},{\mathbf1},{\mathbf1},-\frac{1}{2},\pm1,-\frac{1}{2}\right)$
\\
\midrule
$S+b_2+x+z_2+T_1+T_3$&
$4\times\left(\mathbf{1},{\mathbf1},{\mathbf8},-\frac{1}{2},0,-\frac{1}{2}\right)$\\
$S+b_3+T_2$&
$4\times\left(\overline{\mathbf{16}},{\mathbf1},{\mathbf1},0,0,-\frac{1}{2}\right)$\\
$S+b_3+x+T_1$&
$8\times\left(\mathbf{1},{\mathbf1},{\mathbf1},\pm\frac{1}{2},\pm\frac{1}{2},0\right)$\\
$S+b_3+x+T_2$&
$4\times\left(\mathbf{1},{\mathbf1},{\mathbf8},+\frac{1}{2},-\frac{1}{2},0\right)$\\
$S+b_3+x+z_2+T_1+T_2$&
$4\times\left(\mathbf{1},{\mathbf1},{\mathbf8},-\frac{1}{2},-\frac{1}{2},0\right)$\\
\bottomrule
\end{tabular}
\caption{\it Spectrum of massless fermionic matter and quantum numbers under the various gauge group factors. Sectors in the first column are labeled according to the conventions of the fermionic construction for the basis vectors, as given in \eqref{basis}.}\label{TableSpectrFerm}
\end{table*}


By construction, at the generic point, the model satisfies the super no-scale structure condition $n_B=n_F$. However at the fermionic point, the above expansion reveals that its  physical massless spectrum has $n_B-n_F=+40$, and a naive argument based only on  the  abundance of massless bosons would have already ruled it out, as it would have predicted a puddle-shaped potential with a minimum at negative values. After all, one might think that this is much worse than the case of model B, where the abundance of massless physical bosons ($n_B-n_F=+8$) was much softer. Model C is a striking example where the effect of unphysical tachyons and non level-matched massive states completely alters this naive expectation, as outlined in the previous section.

Indeed, rearranging \eqref{Zmeta1} into an expansion in terms of $q_{\rm r}$,
\begin{align}
	\begin{split}
		Z_{(C)}&=\frac{2 q_{\rm i}}{q_{\rm r}}-\frac{16 q_{\rm i}}{\sqrt{q_{\rm r}}}+\left(40+64 q_{\rm i}+56 q_{\rm i}^2\right)+\left(224+\frac{6912}{q_{\rm i}}+768 q_{\rm i}-672 q_{\rm i}^2\right) \sqrt{q_{\rm r}}\\
		&+\left(14336+\frac{9216}{q_{\rm i}^2}+\frac{118656}{q_{\rm i}}-10144 q_{\rm i}+3072 q_{\rm i}^2+792 q_{\rm i}^3\right) q_{\rm r}  \\
		&+\left(-203776+\frac{934400}{q_{\rm i}^2}+\frac{498224}{q_{\rm i}}-39744 q_{\rm i}+12800 q_{\rm i}^2-10128 q_{\rm i}^3\right) q_{\rm r}^{3/2} +\ldots\,,
	\end{split}
\end{align}
and using the model-independent values for the one-loop potential integrals $I_{n,m}^1$ and $I_{n,m}^2$ given in Tables \ref{tableI1} and \ref{tableI2}, we may estimate the contributions of every mass level $n$ to the one-loop potential and verify that the potential exhibits a positive value.
\begin{table*}\centering
\ra{1.3}
\begin{tabular}{@{}r|rr@{}}\toprule
$n $ & Model  & C   \\ \midrule
$-1$&&  24.4  \\
$-\frac{1}{2}$ &&  $-9.87$  \\
0 &&  $-20.5$ \\
$\frac{1}{2}$ &&  $10.6$  \\
1 &&  $4.04$  \\
$\frac{3}{2}$ &&  $2.73$  \\ \midrule
 Total && +11.4 \\ \bottomrule
\end{tabular}
\caption{{\it Contributions to the rescaled one-loop potential, $2(2\pi)^4 V_{\rm 1-loop}$, arranged according to energy levels for model C. At each level $n$, the cumulative contribution of level-matched as well as non level-matched states is displayed.}}\label{tableNumC}
\end{table*}

A summary of the contributions to the energy budget of $2(2\pi)^4 V_{\rm one-loop}$ arranged per mass level is presented in Table \ref{tableNumC}.  Unphysical tachyons have identical contributions as with model A. As one might expect, the massless level $n=0$ is dominated by level-matched states, which contribute $\sim -22.0$ out of a total $\sim -20.5$. This is due to the relatively large abundance of massless physical bosons. If we were to consider only the lowest mass levels $n\leq 0$, we would have concluded that the rescaled potential is negative at the fermionic point, with value $\sim - 6.00$. The situation changes drastically, however, as soon as one considers the first massive excitations, in particular $n=1/2$. A careful analysis reveals that the unphysical massless states $6912 \sqrt{q_{\rm r}}/q_{\rm i}$ with conformal weights $(-\frac{1}{4},+\frac{3}{4})$ are responsible for the contribution $\sim +11.2$, which effectively brings the potential back to positive values. Similarly, a considerable positive contribution $\sim+10.0$ arises from the non level-matched states $118656 q_{\rm r}/q_{\rm i}$ occurring at $n=1$ with conformal weights $(0,1)$, which reinforces further the positivity of $V_{\rm one-loop}$. Both the positivity of the potential and its numerical value remain essentially unchanged if one keeps increasing the order of truncation to higher masses, and the validity of these statements has been checked to very high levels.

In order to study the precise form of the one-loop potential, we need to define the model at the generic point in the perturbative moduli space, by rewriting it in its orbifold representation and then marginally deform it. As with all models in the class under consideration, the form of the partition function is always the same and was given in \eqref{orbifoldpartitionf}, with the only model dependence entering the modular covariant phase $\Phi$. For model C, it is given by
\begin{align}
	\begin{split}
	\Phi =& ab+k\ell+\rho\sigma\\
		& +ag_2+bh_2+h_2 g_2\\
		& +kG+\ell H+HG\\
		& +\rho G+\sigma H+HG\\
		& +H_1(b+\sigma)+G_1(a+\rho)\\
		& +H_2 \sigma+G_2\rho+H_2 G_2\\
		& +H_3 (\ell+\sigma)+G_3(k+\rho)\\
		& +H_1 g_2+ G_1 h_2\\
		& +H_3 g_2+G_3 h_2\\
		& +H_3 G+G_3 H\\
		& +H_3 G_2+G_3 H_2 \,.
	\end{split}\label{phaseC}
\end{align}
Similarly to models A and B, also model C shares the precise same generic characteristics discussed in section \ref{examplemodel}. In particular, the gravitino mass, gauge group, residual T-duality group and the fact that non-vanishing contributions to the vacuum energy only arise from $h_1=g_1=0$, are precisely the same as the ones given in section \ref{examplemodel} for model A and shall not be repeated here.

The  formal definition of the model as a $T^6 / (\mathbb{Z}_2)^6$ orbifold presents no difficulty and follows immediately from an inspection of the partition function \eqref{orbifoldpartitionf} and the specific choice of modular covariant phase \eqref{phaseC}. The notation is identical to the one employed in sections \ref{examplemodel} and \ref{counterexamplemodel}, with   the six $\mathbb Z_2$ factors being associated with their corresponding boundary condition parameters as in eq. \eqref{z6param}. The action on the worldsheet degrees of freedom reads
\begin{align}
	\begin{split}
		& \mathbb Z_2^{(1)} \ : \ X^{1,2,5,6}\to - X^{1,2,5,6}\\
		& \mathbb Z_2^{(2)} \ : \ X^{3,4,5,6}\to - X^{3,4,5,6}\\
		& \mathbb Z_2^{(3)} \ : \ (-1)^{F_{\rm s.t.}+F_2}\,\delta_1 \quad,\ \delta_1: \{ X_1\to X_1+\pi R_1\} \\
		& \mathbb Z_2^{(4)} \ : \ (-1)^{F_2}\,\delta_3 \quad,\ \delta_3: \{ X_3\to X_3+\pi R_3\} \\
		& \mathbb Z_2^{(5)} \ : \ (-1)^{F_1+F_2}\,\delta_5 \quad,\ \delta_5:\{ X^5\to X^5+\pi R_5\} \\
		& \mathbb Z_2^{(6)} \ : \ (-1)^{F_1}\,r \quad,\ r:\{\bar\phi^{5,6,7,8}\to -\bar\phi^{5,6,7,8}\} \,,
	\end{split}\label{orbdefC}
\end{align}
together with the non-trivial discrete torsion choice
\begin{align}
	\epsilon(2,3),\ \epsilon(2,5),\ \epsilon(4,5),\ \epsilon(5,6)\,.
\end{align}

We are now ready to investigate the form of the one-loop potential as a function of the Scherk-Schwarz moduli $(T,U)$ associated to the first $T^2$. Similarly to our treatment of models A and B, we consider the simplified case where all moduli are kept fixed at their fermionic values as in \eqref{modulideformation} and consider only deformations with respect to the toroidal volume $T_2$ and shape $U_2$. The procedure for simplifying the partition function and, hence, the integrand of \eqref{potentialdef} is similar to the one employed in sections \ref{examplemodel} and \ref{counterexamplemodel}. The integrand is then cast in the generic simplified form of \eqref{Zsimplified}, with the phase $\hat{\Phi}$ corresponding to model C  given by
\begin{align}
	\begin{split}
	\hat\Phi  &=  k\ell + (kG+\ell H+HG)\\
			&+ H_1(g_2+\sigma)+G_1(h_2+\rho)\\
			&+ \gamma_2(G+g_2+\ell+\delta_3) + \delta_2(H+h_2+k+\gamma_3)+\gamma_2\delta_2\\
			&+ \gamma_3\sigma + \delta_3 \rho\\
			&+ h_2\sigma+g_2 \rho+h_2 g_2\\
			&+ \ell \rho + k \sigma\\
			& +g_2(1+H_1+k+h_2) \,.
	\end{split}
\end{align}
\begin{figure}[t]
	\centering
    \includegraphics[width=\textwidth]{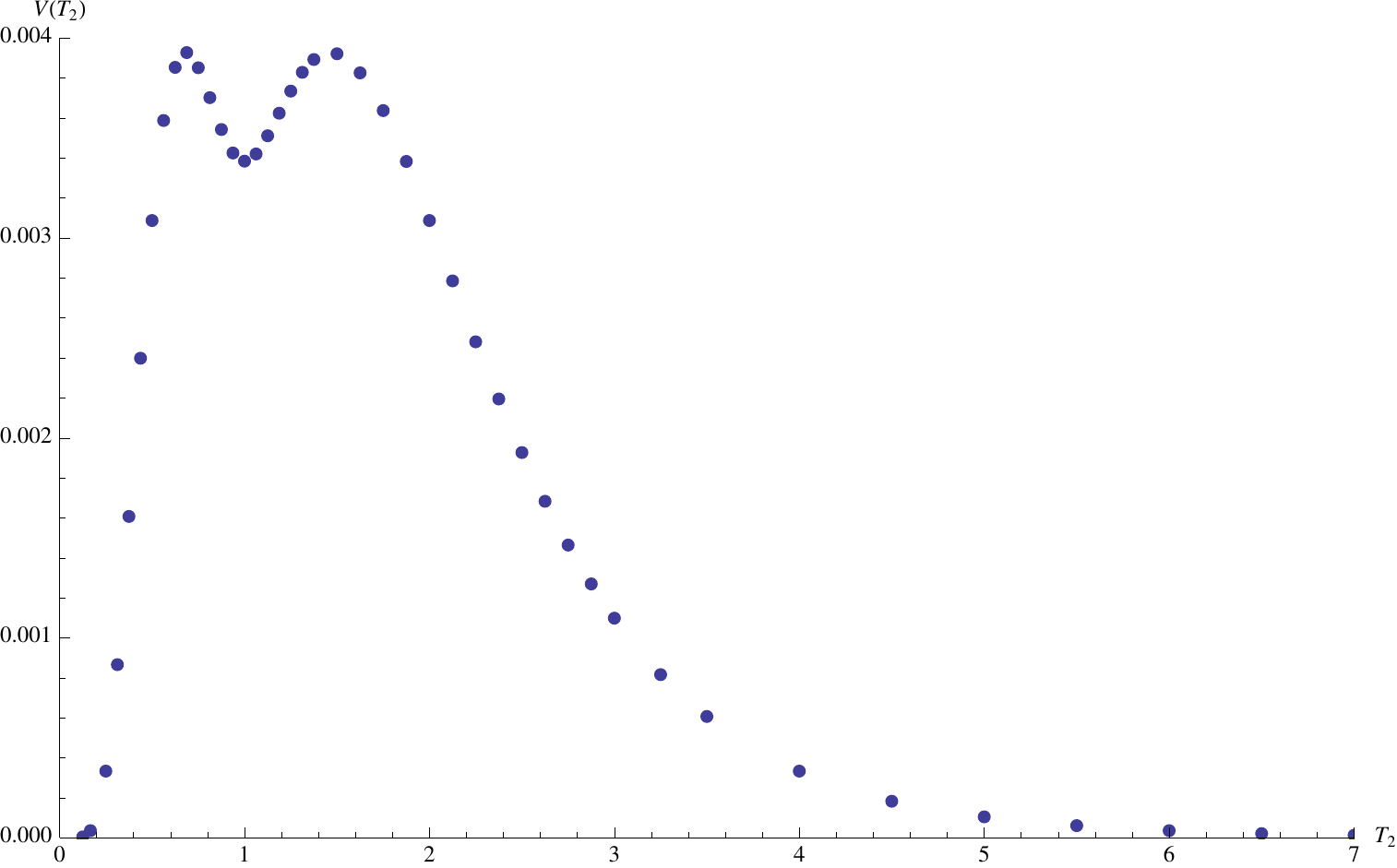}
    \caption{{\it The  one-loop potential as a function of $T_2$ for  Model C, obtained by direct numerical evaluation of the integral  \eqref{potentialdef} without unfolding.}}\label{metaFig}
\end{figure}

As in models A and B, from this simplified form one may extract the asymptotic behaviour \eqref{AsymptoticExpr} of the one-loop integral. Here, the situation closely resembles that of model B, in that it exhibits the super no-scale structure resulting from the condition $n_B=n_F$ at the generic point in the $(T,U)$ moduli space, which causes the power-law behaviour  in the first line of \eqref{AsymptoticExpr} to be absent. For large $T_2\gg 1$, one verifies that the potential is indeed exponentially suppressed and this satisfies our requirement of having a small cosmological constant at large volumes. This was the case also for model B. Unfortunately, the latter model exhibited a minimum at the fermionic point and the corresponding potential had the form of a puddle, dynamically stabilising the volume $T_2$ at the fermionic point and at a huge negative value for the cosmological constant.

In the present case of model C, the situation is drastically different, as outlined in the beginning of the section. This is so because the unphysical tachyons, and massive non level-matched states conspire to overcome the negative contributions caused by the excess of 40 physical massless bosons, and lead to a positive value of the cosmological constant at the fermionic point. The precise form of the one-loop potential $V_{\rm one-loop}$ as a function of the Scherk-Schwarz volume is plotted in Figure \ref{metaFig}, as obtained by direct numerical integration without unfolding. Perhaps not surprisingly, the potential has the form of a local minimum at positive values for $V_{\rm one-loop}$. The presence of this metastable structure can be explained by the fact that, as one deforms away from the fermionic point, the abundance of 40 extra massless bosons must be eliminated in order to reach the super no-scale structure at the generic point. Therefore, as the excess bosons acquire a mass, the contribution of the unphysical tachyons and the non level-matched massive modes responsible for the positivity of the potential becomes even more dominant.

This appears to be precisely the situation we were aiming for. One could imagine a scenario in which the theory starts with the torus volume stabilised at the false vacuum $T_2=1$ and then subsequently decays towards the true vacuum in the regime of large volume and low SUSY breaking scale, while simultaneously suppressing the value of the cosmological constant. 

It turns out that the story is somewhat more intricate. The metastable\footnote{Constructions of metastable vacua in type I string theory were discussed in \cite{Angelantonj:2007ts}.} behaviour with $T_2$ initially stabilised at the fermionic point immediately poses questions concerning the classical stability of the theory with respect to other moduli. Indeed, as we have already mentioned in previous sections, the moment the supersymmetry breaking parameter $T_2$ lies sufficiently close to the string scale one has to worry about the possibility of some BPS states crossing the massless barrier and becoming tachyonic in some region in the $T,U\sim 1$ parameter space.

Tachyonic states may in principle appear as scalars arising from the $H_1=1$  sector. In the case under consideration, a careful analysis shows that the first potentially tachyonic states have the mass formula 
\begin{align}
	M_{\rm BPS}^2 = \frac{1}{2}\left( T_2 + \frac{1}{T_2}\right)\left( U_2+\frac{1}{4U_2}-\left|U_2-\frac{1}{4U_2}\right|\right)-1\,,
	\label{BPStach}
\end{align}
which exhibits the invariance under the $\Gamma^1(2)_T$ T-duality transformation $T_2\to 1/T_2$, and under $U_2\to 1/(4U_2)$. Indeed, an analysis of the shifted lattice reveals that the transformation
\begin{align}
	\begin{pmatrix}
										1 & -1 \\
										2 & -1 
								\end{pmatrix} \in \Gamma_0(2)_U \quad\leftrightarrow\quad U\to \frac{U-1}{2U-1} 
\end{align}
is a symmetry of the theory such that, when acting on $U=\frac{1}{2}+iU_2$, it effectively transforms $U_2\to 1/(4U_2)$. Hence, the allowed region in the $(T_2,U_2)$ parameter space that guarantees the tree-level stability of the theory is
\begin{align}
	\left(T_2+\frac{1}{T_2}\right)^{-1} \leq U_2 \leq \frac{1}{4}\left(T_2+\frac{1}{T_2}\right) \,.
	\label{allreg}
\end{align}
The bounds are saturated precisely at those points where the BPS states \eqref{BPStach} cross the massless barrier, beyond which they become tachyonic. 

\begin{figure}[t]
	\centering
    \includegraphics[width=\textwidth]{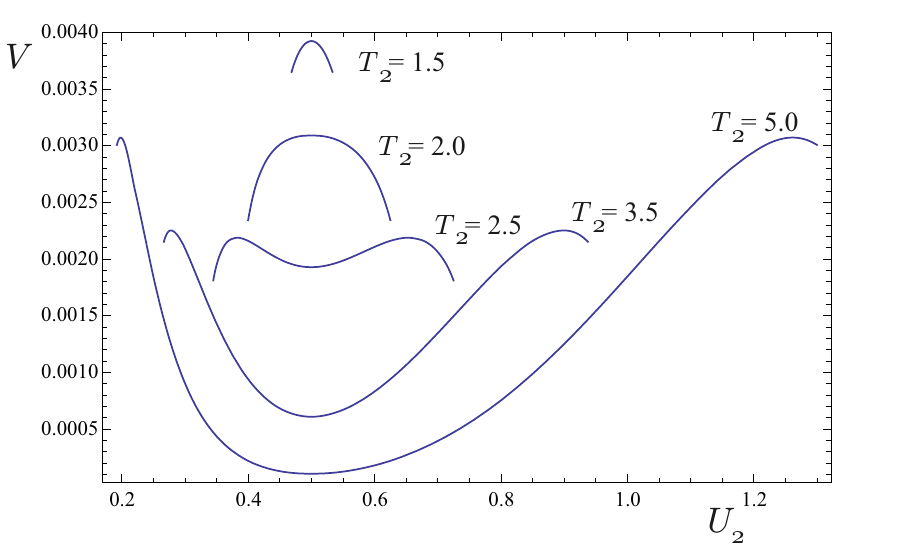}
    \caption{{\it The one-loop potential for Model C as a function of $U_2$, plotted for different values of the volume $T_2$. }}\label{U2plot}
\end{figure}

\begin{figure}[t]
	\centering
    \includegraphics[width=\textwidth]{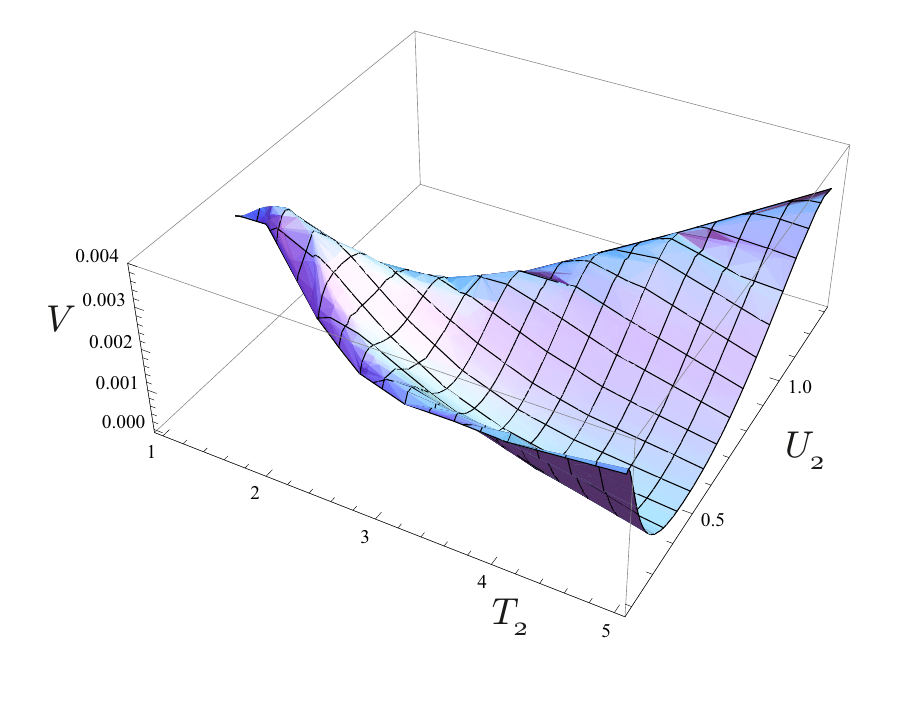}
    \caption{{\it Numerical reconstruction of the one-loop potential for Model C as a smooth function of $T_2$ and $U_2$ within the allowed parameter space defined by eq. \ref{allreg}.}}
    \label{3DPLOTb}
\end{figure}

Since  tachyons are necessarily bosons that become massless at these points, one would naively expect that for a given fixed value of $T_2>1$, the one-loop potential exhibits a local maximum at $U_2=1/2$ which destabilises the theory, leading it straight into the tachyonic regime. Fortunately, this is not the case, for the same reasons discussed in the beginning of this section. Namely, the contributions of non-level matched states or even massive states can be highly non-trivial in the vicinity of self-dual points. The results of our numerical analysis of the one-loop potential at high order precision in the $q_{\rm r}$ expansion are summarised in Figures \ref{U2plot}-\ref{3DPLOTb}. At least for values of $T_2 \gtrsim 2.20$, these contributions actually cause the potential to exhibit precisely the opposite behaviour: they generate an attractor that stabilises $U_2$ at its fermionic value $U_2=1/2$, while maintaining the rollout of $T_2$ to the large volume regime and, therefore, the dynamical consistency of the model is guaranteed.



\section{Conclusions}\label{conclusions}

The possibility of constructing viable heterotic theories with spontaneous supersymmetry breaking via Scherk-Schwarz fluxes is a very appealing one. Some of the major problems plaguing supersymmetric string theories, such as the stabilisation of moduli and the degeneracy of vacua, are lifted as soon as supersymmetry is broken. In particular, quantities of phenomenological interest which remain undetermined in a supersymmetric setup, such as the gravitino mass scale, may be dynamically fixed by radiative corrections to the scalar potential. 

One might imagine an idealised scenario in which the rich structure of string radiative corrections and a deep understanding of contributions to the effective potential could even furnish us with a dynamical mechanism able to explain the number of non-compact spacetime dimensions of our low energy world. Although our present conceptual and technical understanding of strings in non-supersymmetric setups is still very limited compared to supersymmetric ones, it is an interesting and important problem to analyse the implications of such theories.

In practice, taming radiative corrections in the absence of supersymmetry appears to be a rather delicate task. On the one hand, for such theories to be viable, they need to be supplemented with a mechanism that dynamically secures their classical stability against the presence of tachyonic modes. On the other hand, a non-vanishing value of the vacuum energy already at one loop signals a dilaton tadpole that necessitates a proper treatment of the back-reaction problem.

In this work, we propose that both issues may be to some extent addressed in one stroke by constructing super no-scale models which, at least in a wide region of parameter space, are dynamically stable and which naturally select supersymmetry breaking at low scales $m_{3/2}\ll M_{\rm Planck}$, while maintaining a controllable exponentially suppressed value for the cosmological constant.

To this end, we exploited the equivalence between fermionic and orbifold constructions at special points in moduli space, in order to scan a random sample of $10^6$ models subject to certain criteria, such as the presence of chiral matter and an observable ${\rm SO}(10)$ gauge group factor. Working in the interplay between the two formulations, it was possible to study the contributions of various states to the one-loop effective potential and derive a set of conditions \eqref{conditionRefined} that guarantee its positivity. 

Our central observation is that massive and even non-level matched states  play a significant role in determining the morphology of the effective potential around special self-dual points. This result, although counter-intuitive from a field theoretic perspective, was central to our analysis and resulted in the construction of the explicit example `Model C' defined in \eqref{orbdefC} that illustrates the desired behaviour for the one-loop potential.

Of course, our present analysis is only a first step in this very interesting direction and there are several open questions that deserve future investigation. On the one hand, the specific construction of Model C is by no means unique but only a particular solution to our computer-aided scan in a random sample of $10^6$ models. It is plausible that, by extending the chosen basis \eqref{basis} \emph{e.g.} to further break the SO(10) group, one might obtain similar realisations with Standard-Model like matter content. Another question concerns the shape of the potential with respect to other moduli that, for reasons of simplicity, were held fixed in this work. Although the latter are not expected to significantly modify our main results, it is still interesting to analyse their role and stability properties. 

On the other hand, our present results serve to illustrate some of the richness of the effective potential as one probes regions close to the string scale. It is possible that, aside from the fermionic point, also other special points may play a significant role in determining the shape of the effective potential, perhaps opening the possibility for constructing mestable vacua similar to Model C, but with the false vacuum being protected against the development of tachyonic instabilities for all relevant moduli.

Finally, we wish to mention another related open question, concerning the fate of the running of couplings in chiral non-supersymmetric models. It is by now fairly well understood that the presence of chirality is inherently linked to the decompactification problem \cite{Kiritsis:1996xd} whenever the volume of the internal space becomes sufficiently larger than the string scale. Although some proposals have been put forward in the literature, effectively securing the theory from the strong coupling regime \cite{Faraggi:2014eoa}, they were constrained only to the non-chiral case. It would be interesting to re-evaluate this problem and see whether other perturbative or non-perturbative effects could produce a remedy.



\section*{Acknowledgements}

We would like to thank C.~Angelantonj, C.~Kounnas, H.~Partouche, N.~Rompotis, N.~Tetradis and N.~Toumbas for useful discussions and I.F. wishes to thank the University of Ioannina for its warm hospitality, during the early stages of this work.

\bibliographystyle{utphys}
\providecommand{\href}[2]{#2}\begingroup\raggedright\endgroup

\end{document}